\documentclass[a4paper,fleqn]{cas-dc}

\usepackage[authoryear]{natbib}
\usepackage{aas_macros}
\begin{document}
\let\WriteBookmarks\relax
\def\floatpagepagefraction{1}
\def\textpagefraction{.001}
\shorttitle{A review of the DIM}
\shortauthors{J.M. Hameury}

\title [mode = title]{A review of the disc instability model for dwarf novae, soft X-ray transients and related objects}

\author[1]{J.M. Hameury}[orcid=0000-0002-6412-0103]
\ead{jean-marie.hameury@astro.unistra.fr}
\address[1]{Observatoire Astronomique de Strasbourg}

\begin{abstract}
I review the basics of the disc instability model (DIM) for dwarf novae and soft-X-ray transients and its  most recent developments, as well as the current limitations of the model, focusing on the dwarf nova case. Although the DIM uses the Shakura-Sunyaev prescription for angular momentum transport, which we know now to be at best inaccurate, it is surprisingly efficient in reproducing the outbursts of dwarf novae and soft X-ray transients, provided that some ingredients, such as irradiation of the accretion disc and of the donor star, mass transfer variations, truncation of the inner disc, etc., are added to the basic  model. As recently realized, taking into account the existence of winds and outflows and of the torque they exert on the accretion disc may significantly impact the model. I also discuss the origin of the superoutbursts that are probably due to a combination of variations of the mass transfer rate and of a tidal instability. I finally mention a number of unsolved problems  and caveats, among which the most embarrassing one is the modelling of the low state. Despite significant progresses in the past few years both on our understanding of angular momentum transport, the DIM is still needed for understanding transient systems.
\end{abstract}


\maketitle
\sloppy

\section{Introduction}
Dwarf novae (DNe) are cataclysmic variables (CV) that undergo regular outbursts \citep[see][for an encyclopaedic review]{w03}, in contrast to novalike systems that are also CVs, but do not show such outbursts. Both classes have been extensively observed for more than one hundred years, in particular by amateur astronomers, and continuous observations of many systems over a long period of time have been made available. Transient systems are important because they allow to probe the fundamental properties of accretion discs; it is for example well known that the observational appearance of steady accretion discs do not depend on the mechanisms that transport angular momentum, provided that they can be described as viscosity \citep{fkr02}. 

It has been known for long that molecular viscosity is by far too small to account for the observed variability time scales of accretion flows, if they are related to the viscous time; moreover, unless the disc mass builds up to unrealistic, large values, the mass accretion rate would be vanishingly small. Other processes involving turbulence and/or magnetic fields must therefore play a role. \citet{ss73} parametrized our ignorance using a simple prescription that proved to be surprisingly efficient. It is now widely believed that the anomalous viscosity in accretion discs is due to the magnetorotational instability (MRI) \citep{bh91}, first discovered by \citet{v59} and \citet{c60}, although hydrodynamic turbulence has also been considered \citep[see e.g.][and references therein]{fl19}.

Spiral shocks \citep{smi87,jsz16} or waves \citep{kbk14} may also transport angular momentum; spiral patterns have been observed in several systems using Doppler tomography techniques (\citealt{shh97,mh88}; see also \citealt{ms16} for a recent review). However, their efficiency when the Mach numbers are high as expected in cataclysmic variables could be vanishingly small. 

Whatever the nature of the mechanism transporting angular momentum, it is widely accepted that dwarf nova outbursts are caused by a thermal instability developing in the accretion disc when the central temperature is of the order of the ionization temperature of hydrogen and the opacities are strongly dependent on temperature. This idea encapsulated in the so-call disc instability model (DIM) has been developed for almost 40 years beginning with \citet{mm81,s82,cgw82,flp83,mo83}, to name a few of the initial authors.

It was then suggested that the transient nature of some low-mass X-ray binaries (LMXBs), the so-called soft X-ray transients, could be explained by the same model \citep{vpv84,cwg85}, although significant modifications had to be introduced \citep{v96}. 

Extension of the DIM to young stars \citep[see e.g.][for FU Orionis stars]{bl94} and to active galactic nuclei \citep{jcs04,hvl09} have also been considered, but were not as successful, and will not be discussed here.

For a while, it was hoped that one could infer the properties and characteristics of angular momentum transport in accretion discs by comparing the predictions of the DIM with observations. It was then realized that, in order to account for a number of observational properties of DNe, additional ingredients such as mass transfer variations, truncation of the inner disc, irradiation of the accretion disc, winds and outflows, etc. had to be added to the model, introducing additional free parameters that are not well constrained; but generally speaking, the model agrees fairly well with observations. At the same time, much progress was made on the MRI, but numerical simulations have for long had difficulties in accounting for an efficient enough angular momentum transport, in particular when no net magnetic flux is assumed. We are therefore facing an interesting contradiction: the DIM explains well the observed properties of dwarf novae, but its physical foundations (that angular momentum transport can be described by a Shakura-Sunyaev type viscosity) are weak, whereas the MRI, that is developed on strong theoretical grounds, does not produce as good results as one would have hoped. 

It has been almost twenty years since the last thorough and complete review of \citet{l01}. During this period, the basics of the DIM have not changed much, but a lot of progress has been made on related items such as angular momentum transport in accretion discs; new observations, in particular by \emph{Kepler} have brought new data with exquisite quality. Additional ingredients to the DIM have also been added in order to account for these new observations. In this context, it is important to understand why the DIM is still needed to understand transient systems and what its limitations are.

In this paper, I shall deal with both dwarf novae and soft X-ray transients, with a strong focus on dwarf novae. Section 2 briefly describes the categories and sub-categories of objects that will be discussed in this review. I summarize in section 3 the main ingredients of the thermal / viscous instability model for dwarf novae; the reader is invited to refer to \citet{l01} for more details. I then discuss the various extensions of the model that are needed to account for observations in section 4, and present a few current problems with the DIM in section 5.

\section{The zoo of dwarf nova type outbursts} \label{sect:zoo}
As we shall see, the DIM can produce a large variety of outbursts, thereby accounting for the zoo of observed eruptive systems. Among cataclysmic variables, dwarf novae are divided in several subclasses, labelled according to a supposedly representative member, according to the outbursts properties. I summarize here the most important ones \citep[see][for a detailed description of these subclasses]{w03}
\begin{itemize}
\item U Gem type systems are the classical dwarf novae; U Gem undergoes $\sim 5$ mag. outbursts in the optical lasting typically for 1-2 weeks and recurring every 3 months.
\item Z Cam systems show long standstill periods (a few months) interrupting a characteristic sequence of classical dwarf nova outbursts. During standstills, the system is slightly fainter (typically 1 mag) than at the outburst maximum.
\item SU UMa systems show regularly long outbursts, called superoutbursts, lasting 5-10 times longer than normal outbursts and occurring every few normal outbursts. These will be specifically discussed in Sect. \ref{sec:tti} and \ref{sec:suuma}. This subclass is further divided into several sub-subclasses, including the WZ Sge systems, that show only superoutbursts, and the ER UMa systems that are characterized by their extremely high outburst frequency and short (19--48 d) supercycles.
\end{itemize}
This classification intersects the magnetic/non-magnetic classification of CVs. Whereas no CV in which the white dwarf magnetic field is strong enough to synchronize its rotation with the orbital period (the so-called AM Her systems or polars) shows outbursts, some of the less magnetized systems (the intermediate polars) do show dwarf nova outbursts. It also intersects the classification of CVs according to the composition of the secondary: some systems in which the secondary is helium-rich, the so-called AM CVn systems do show outbursts.

Finally, some systems, called VY Scl stars, do not show dwarf nova outbursts and are therefore classified as novalike variables, but fade at random intervals to low-luminosity states during which they can be as faint as dwarf novae in quiescence; they are sometimes called anti dwarf novae. On rare occasions, and in very few systems, outbursts may occur during low states \citep{sp98}.

Symbiotic stars are binary systems closely related to CVs, in which a hot primary (usually a white dwarf, but a neutron star has been found in a few systems) accretes from a giant secondary that may or may not fill its Roche lobe. Their behavior has some similarities with CVs, including nova eruptions, and symbiotics might in principle also be subject to dwarf nova instabilities.

Finally, as mentioned earlier, low-mass X-ray binaries may show a dwarf nova type behavior, in the case of soft X-ray transients. The outbursts last typically for weeks, and their recurrence time ranges from years (in the case of neutron star binaries) to decades (black hole binaries) \citep[see e.g.][for reviews of the observations]{ts96,csl97,yy15}. 

\section{Basics of the disc instability model}
A major assumption of the model is that angular momentum transport is due to a local mechanism and is accompanied by local energy release, so that it can be described by some ''effective'' viscosity. 

If the accretion disc is geometrically thin, quantities can be integrated in the vertical direction, and the orbital motion is Keplerian (except in the boundary layer, not considered here). As one also assumes axisymmetry, the problem is one dimensional. In cylindrical coordinates, the equations for mass and angular momentum conservation can be written as \citep[see e.g.][]{l01}:
\begin{equation}
\frac{\partial \Sigma}{\partial t} = - \frac{1}{r} \frac{\partial}{\partial r} (r \Sigma v_r) + \frac{1}{2 \pi r} \frac{\partial \dot{M}_{\rm ext}}{\partial r}
\label{eq:consm}
\end{equation}
and
\begin{eqnarray}
j\frac{\partial \Sigma}{\partial t} & = & - \frac{1}{r} \frac{\partial}{\partial r} (r \Sigma j v_{\rm r})  + \frac{1}{r} \frac{\partial}{\partial r} \left(- \frac{3}{2} r^2 \Sigma \nu \Omega_{\rm K} \right) + \nonumber \\ & & \frac{j_{\rm k}}{2 \pi r} \frac{\partial \dot{M}_{\rm ext}}{\partial r} - \frac{1}{2 \pi r} T_{\rm tid}(r),
\label{eq:consj}
\end{eqnarray}
where $\Sigma$ is the surface column density, $\dot{M}_{\rm ext}(r)$ the rate at which mass is incorporated into the disc at point $r$, $v_{\rm r}$ the radial velocity in the disc, $j = (GM_1r)^{1/2}$ is the specific angular momentum of material at radius $r$ in the disc, $\Omega_K=(GM_1/r^3)^{1/2}$ is the Keplerian angular velocity ($M_1$ being the mass of the accreting object), $\nu$ is the kinematic viscosity coefficient, and $j_{\rm k}$ the specific angular momentum of the material transferred from the secondary. $T_{\rm tid}$ is the torque due to tidal forces, for which several prescriptions have been proposed; one often uses that of \citet{s84}, derived from the determination of tidal torques by \citet{pp77}):
\begin{equation}
T_{\rm tid} = c \omega r \nu \Sigma \left( \frac{r}{a} \right)^{5},
\label{eq:defT}
\end{equation}
where $\omega$ is the angular velocity of the binary orbital motion, $c$ is a numerical coefficient taken so as to give a stationary (or time averaged) disc radius equal to some chosen value, and $a$ is the binary orbital separation. This prescription enables significant variations of the outer radius of the disc, but does not ensure that the disc does not extend beyond the tidal truncation radius. \citet{vh08} assumed that the tidal torques are negligibly small as long as the disc stays within the tidal truncation radius $r_{\rm tid}$ and become arbitrarily large for $r>r_{\rm tid}$, thereby ensuring that the radius does not exceed $r_{\rm tid}$, or equivalently, that
\begin{equation}
T_{\rm tid} = c \omega r \nu \Sigma e^{K(r-r_{\rm tid})/a}
\end{equation}
where $K$ is a constant much larger than unity \citep{hl16}.

The energy conservation equation can be written as:
\begin{equation}
\frac{\partial T_{\rm c}}{\partial t} = \frac{ 2 (Q^ + -Q^- + J)}{C_P \Sigma} - \frac{\Re T_{\rm c}}{\mu C_P} \frac{1}{r} \frac{\partial (r v_r)}{\partial r} - v_r \frac{\partial T_{\rm c}}{\partial r},
\label{eq:heat}
\end{equation}
where $Q^+$ and $Q^-$ are the surface heating and cooling rates respectively; $Q^+=(9/8) \nu \Sigma \Omega_{\rm K}^2$ and $Q^- = \sigma T_{\rm eff}^4$, $T_{\rm eff}$ being the effective temperature \citep[e.g.][]{c93}).  $T_{\rm eff}$ must be provided as an input, and can be calculated as a function of $\Sigma$, $T_{\rm c}$ and $r$ as will be shown later in Sect. \ref{sec:vert_str}. The term $J$ accounts for the radial energy flux carried by viscous processes and by radiation. In the simple case where viscosity is due to hydrodynamical turbulence, $J$ can be estimated in the framework of the $\alpha$ parametrization. The flux carried in eddies with characteristic velocity $v_{\rm e}$ and size $l_{\rm e}$, is:
\begin{equation}
F_{\rm e} = C_P \Sigma v_{\rm e} \frac{\partial T_{\rm c}}{\partial r} l_{\rm e} = \frac{3}{2} \nu C_P \Sigma \frac{\partial T_{\rm c}}{\partial r},
\label{eq:fturb}
\end{equation}
in which case
\begin{equation}
J = 1/r \frac{\partial}{\partial r }(r F_{\rm e}).
\end{equation}
Other prescriptions for $J$ exist; they give results very similar to those obtained using Eq.~(\ref{eq:fturb}). The reason for this is that this term is important only in restricted regions where gradients are strong; \citet{mhs99} showed that changing $J$ by one order of magnitude has limited consequences.

The outer boundary conditions are
\begin{eqnarray}
\lefteqn{\dot{M}_{\rm tr} = 2 \pi r \Sigma_0 (\dot{r}_0 - v_{\rm r,0})} \label{eq:bc1} \\
\lefteqn{\dot{M}_{\rm tr} \left[ 1 - \left( \frac{r_{\rm k}}{r_0}\right)^{1/2} \right] = 3 \pi \nu \Sigma_0,}
\label{eq:bc2}
\end{eqnarray}
where the index 0 denotes quantities measured at the outer edge, and $r_{\rm k}$ is the circularization radius, i.e. the radius at which the Keplerian angular momentum is that of the matter lost by the secondary star, and $\dot{r}_0$ is the time derivative of the outer disc radius. Note that the outer disc radius is not known a priori, but is determined by the two sets of equations (\ref{eq:bc1}) and (\ref{eq:bc2}).

In the case of accretion onto a white dwarf or a neutron star, the inflowing matter exerts a torque on the primary that is dissipated in a boundary layer where the angular velocity steeply decreases from the Keplerian to the stellar value and whose radial extend $l$ is small. The boundary layer is not included in the description of the thin disc that therefore ends at the stellar radius plus $l$. One also assumes that the boundary layer does not affect the the thin disc structure, so that the inner boundary conditions is usually taken to be the no-stress condition $\nu \Sigma = 0$, at $r=r_{\rm in} + l \simeq r_{\rm in}$, where $r_{\rm in}$ is the primary radius, which reduces to $\Sigma =0$ \citep{fkr02}. The same no-stress condition applies in the case of accretion onto a black hole. This is set at the inner disc edge, that can be either the radius of the accreting object, or, if magnetic, to the magnetospheric radius (see below). If the primary is magnetized, the disc exerts a torque on the primary, and, in principle, the no-stress condition does not apply. This torque is not simple to evaluate, and, in practice, the condition $\Sigma =0$ is still used, even though in a few cases \citep[see e.g.][]{hl02,hl17}, a different boundary condition was considered. This lead to modifications of the solution close to the inner edge of the disc; in the steady state case, the term $[1-(r_{\rm in}/r)^{1/2}]$ that appears in the standard solution was replaced by 1, and the differences with respect to the no-stress case were significant but not major.

\subsection{Vertical structure and stability}
\label{sec:vert_str}

The disc vertical structure decouples from the radial structure when the disc is geometrically thin, and can be determined by solving the following equations:
\begin{eqnarray}
\lefteqn{\frac{dP}{dz} = -\rho g_{\rm z} = -\rho \Omega_{\rm K}^2 z, }
\label{eq:stra}\\
\lefteqn{\frac{d\ln T}{d  \ln P} = \nabla,} \label{eq:strb}\\
\lefteqn{\frac{dF_{\rm z}}{dz} = \frac{3}{2} \alpha \Omega_{\rm K} P + \frac{dF_t}{dz},} \label{eq:strc}
\end{eqnarray}
where $P$, $\rho$ and $T$ are the pressure, density and temperature respectively, $g_{\rm z} = \Omega_{\rm K}^2 z$ the vertical component of gravity, $F_{\rm z}$ is the vertical energy flux and $\nabla$ the temperature gradient of
the structure. These equations are complemented by an equation of state $P(\rho, T)$ that accounts for partial ionization by solving the Saha equations, and for radiation pressure (that is usually unimportant). They are very similar to the equations describing the radial structure of stars. $dF_t /dz$ is a time-dependent term that describes the departure from thermal equilibrium; one uses the simplifying assumption that this term has the same vertical dependence as the viscous heating term, in which case it can be merged with it by replacing the viscosity parameter $\alpha$ by an effective, unknown value $\alpha_{\rm eff}$. The set of equations (\ref{eq:stra} -- \ref{eq:strc}) can be solved for given radii, surface densities, central temperatures and provide as an output the effective temperature and $\alpha_{\rm eff}$, i.e. the heating and cooling terms that can then be used for solving the radial disc structure. Note that $\alpha_{\rm eff}$ need not be equal to $\alpha$; the equality is enforced only when the radial terms appearing in Eq. \ref{eq:heat} are unimportant, which is the case in steady discs.

\begin{figure}
\includegraphics[width=\columnwidth]{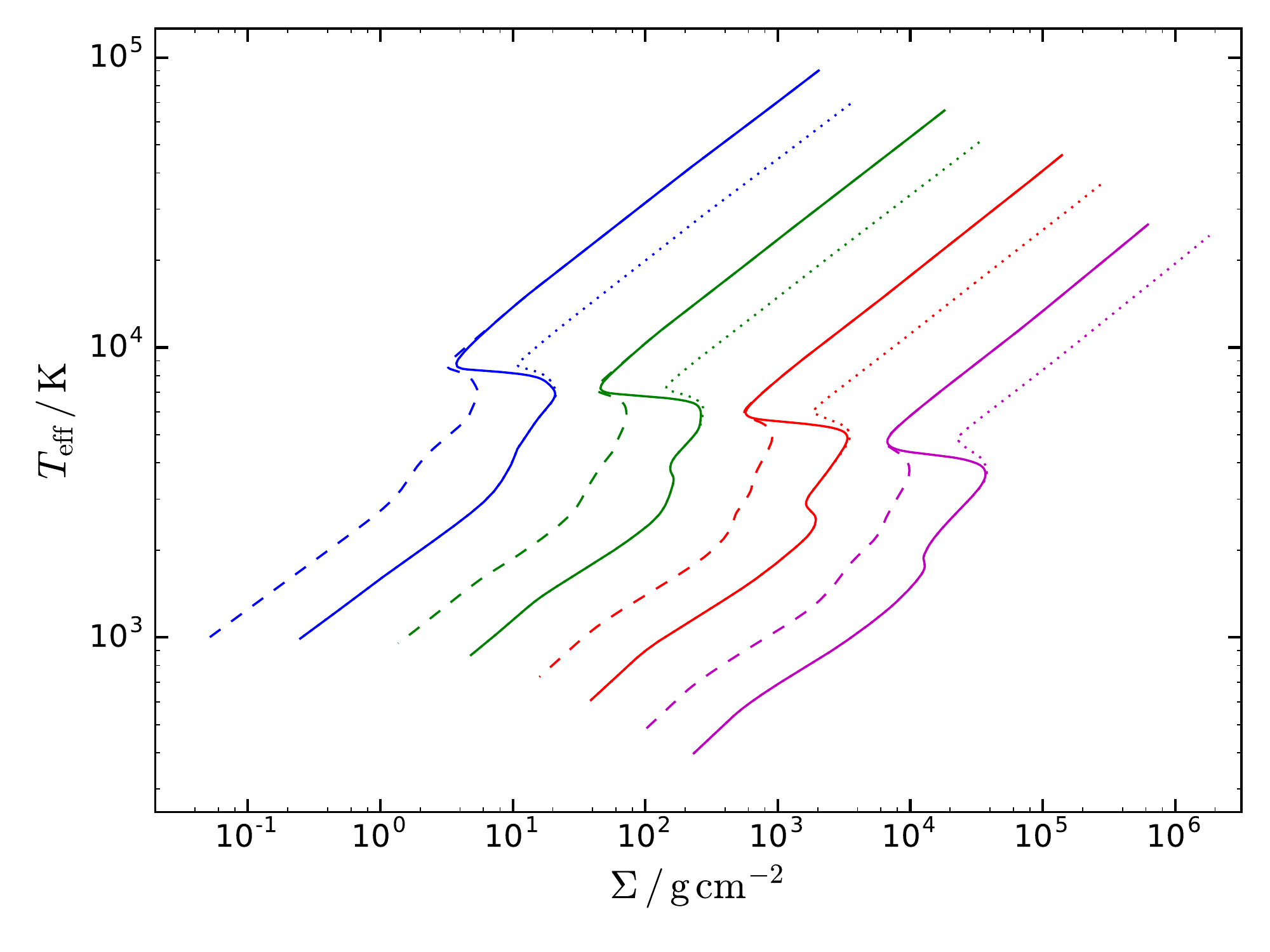}
\caption{Example of $\Sigma - T_{\rm eff}$ S-curves calculated for a 1.35 M$_\odot$ white dwarf at radii  $10^9$, $10^{10}$, $10^{11}$ and $10^{12}$ cm; the dotted and dashed lines correspond to $\alpha$ = 0.01 and 0.1 respectively and the solid line to the interpolation between these two values of $\alpha$ \citep[from][]{bhl18}.}
\label{fig:scurve}
\end{figure}

When one plots the effective temperature (or equivalently, the local accretion rate) as a function of the local surface density, one obtains the famous S-curves that show three branches: a cool branch on which hydrogen is neutral or in molecular form, a hot one where hydrogen is fully ionized, and an intermediate one (see Fig. \ref{fig:scurve}). The hot and cold branches are found to be stable, whereas the intermediate one is both thermally and viscously unstable. However, the stability analysis is local; in order to determine how this instability can propagate throughout the disc, one must solve the radial structure equations (\ref{eq:consm}), (\ref{eq:consj}), and (\ref{eq:heat}) above. 

As discussed below, $\alpha$ is assumed to be different on the hot and on the cold branch; the ratio $\alpha_{\rm h}/\alpha_{\rm c}$ of the hot to cold values has to be typically of order of $5 - 10$. In these conditions, the width of the unstable branch is not determined by the properties of the opacities used, but instead by the change in the $\alpha$ parameter and more precisely by  $\alpha_{\rm h}/\alpha_{\rm c}$.

The upper turning point corresponds to a local accretion rate of \citep{ldk08}:
\begin{equation}
\dot{M}_{\rm crit}^+ = 8.07 \times 10^{15} r_{10}^{2.64}M_1^{-0.89} \; \rm g s^{-1}
\label{eq:mdotcrit+}
\end{equation}
where $r_{10}$ is the radial distance in units of 10$^{10}$ cm, while the lower turning point corresponds to 
\begin{equation}
\dot{M}_{\rm crit}^- = 2.65 \times 10^{15} r_{10}^{2.58}M_1^{-0.85} \; \rm g s^{-1}
\label{eq:mdotcrit-}
\end{equation}
This means that the disc is locally unstable if the local mass accretion rate is in the range $\dot{M}_{\rm crit}^- < \dot{M} < \dot{M}_{\rm crit}^+$. Note that $\dot{M}_{\rm crit}^-$ and  $\dot{M}_{\rm crit}^+$ are independent of $\alpha$; this is due to the fact that the temperature at the turning points are determined by the ionization state of hydrogen. At the critical points, the central temperatures are close to 8250 and 30,000K, and the effective temperatures to 5200 and 6900K, with a weak dependency on mass, radius and $\alpha$. 
The critical surface densities are given by:
\begin{eqnarray}
\lefteqn{\Sigma_{\rm min} = 39.9 \alpha_{0.1}^{-0.80} r_{10}^{1.11}M_1^{-0.37} \; \rm g cm^{-2}} \\
\label{eq:smin}
\lefteqn{\Sigma_{\rm max} = 74.6 \alpha_{0.1}^{-0.83} r_{10}^{1.18}M_1^{-0.40} \; \rm g cm^{-2}}
\label{eq:smax}
\end{eqnarray}
where $\alpha_{0.1}$ is the viscosity parameter normalized to 0.1.

\subsection{Viscosity} 
The extend to which angular momentum transport can be described by the alpha prescription is not well known. Observations however indicate that $\alpha$ is of order of 0.1 or more during outbursts \citep{s99,kpl07,kl12}, and that $\alpha$ is much smaller during quiescence \citep{csw88,csw12}. The estimates of $\alpha$ during outbursts were obtained by comparing the observed relation between the decay time and the orbital period with predictions from the DIM. 
A similar analysis can be performed by comparing the observed and predicted outburst duration and yields similar results. These predictions are robust and almost model-independent, because they basically rely on the assumption that the decay time is driven by the viscous processes in the disc and can be estimated by order of magnitude estimates and simple analytical calculations. These estimates are confirmed and refined by numerical simulations using the DIM. The determination of $\alpha$ during quiescence is based on the recurrence time of outbursts and is somewhat more uncertain, although $\alpha$ must clearly be smaller in quiescence than during outbursts. \citet{mnp19} considered other systems (X-ray binaries, Be stars, FU Ori and protostellar discs) and came to a similar conclusion. 

The standard version of the DIM therefore uses $\alpha_{\rm c} \sim 0.02 - 0.04$ and $\alpha_{\rm h} \sim 0.1 - 0.2$, with an interpolation formula between the hot and cold branches. For example, \citet{hmd98} use:
\begin{eqnarray}
\log (\alpha)& = & \log(\alpha_{\rm c}) + \left[ \log(\alpha_{\rm h})-
\log( \alpha_{\rm c} ) \right] \nonumber \\ & &
\times \left[1+ \left( \frac{2.5 \times 10^4 \; \rm K}{T_{\rm c}} \right)^8
\right]^{-1} .
\label{eq:alpha}
\end{eqnarray}
These values have been somewhat difficult to reconcile with those inferred from the results of numerical simulations of the MRI. We return to this point in Sect. \ref{sec:viscosity}.

\begin{figure}
\includegraphics[width=\columnwidth]{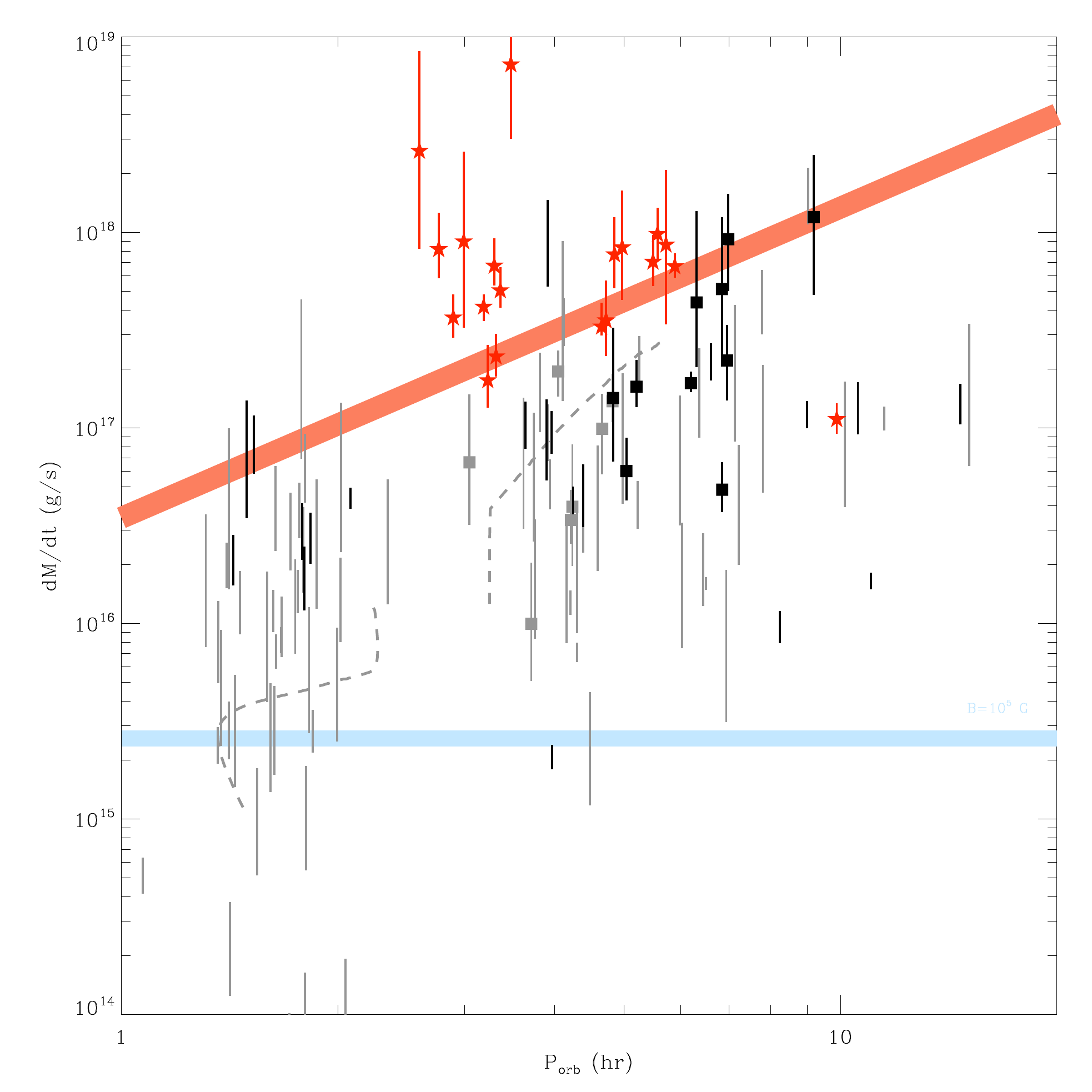}
\caption{Mass transfer rates of cataclysmic variables compared to the stability criterion. Systems above the (red) upper solid line are hot and stable. Systems below the lower (blue) line will be cold, stable discs if the white dwarf magnetic field $B\geq 10^5$\,G. Dashed line is the expected secular mass transfer rate \citep{kbp11}. Square symbols indicate Z Cam type dwarf novae, (red) stars indicate nova-likes. Dwarf novae with $f\geq 0.5$ are shown in black and those with $f<0.5$ are in grey, $f$ being the filling fraction of the light curve, low $f$ values indicating poor time coverage and hence the possibility of having missed outbursts. \citep[from][]{dol18}. Reproduced with permission from Astronomy and Astrophysics \copyright ESO.}
\label{fig:mdot_obs}
\end{figure}

\subsection{Predictions of the DIM and comparison with observations}
According to the DIM, the disc is found to be stable provided that the mass transfer rate from the secondary is either smaller than $\dot{M}_{\rm crit}^-$ everywhere in the disc, or larger than $\dot{M}_{\rm crit}^+$ everywhere in the disc, i.e. $\dot{M}_{\rm tr} < \dot{M}_{\rm crit}^-(r_{\rm in})$ or $\dot{M}_{\rm tr} > \dot{M}_{\rm crit}^+(r_{\rm out})$, where $r_{\rm in}$ and $r_{\rm out}$ are the inner and outer disc radii. The first condition is very difficult to meet in hydrogen dominated disc, unless the mass transfer rate is extremely low, or the inner disc radius is much larger than the radius of a white dwarf (typically a factor 10 or more, see Eq. \ref{eq:mdotcrit-}). \citet{dol18} tested from a sample of about 130 CVs if the above inequalities were satisfied, using the \emph{GAIA} distances to derive the accretion rate. They found that none of the analyzed systems present a challenge to the model. Figure \ref{fig:mdot_obs} indeed shows that dwarf novae are consistent with the unstable region delineated by $\dot{M}_{\rm crit}^+$, with Z Cam systems being at the upper end of the allowed range. Conversely, novalike systems are all above $\dot{M}_{\rm crit}^+$, with the exception of AE Aqr that is an intermediate polar in a propeller state, therefore not well described by their analysis. This general agreement is remarkable, even though the system parameters (masses, inclination) are difficult to estimate and were chosen arbitrarily for a significant fraction of systems shown in Fig. \ref{fig:mdot_obs}, accounting for the large error bars.

\begin{figure}
\includegraphics[width=\columnwidth]{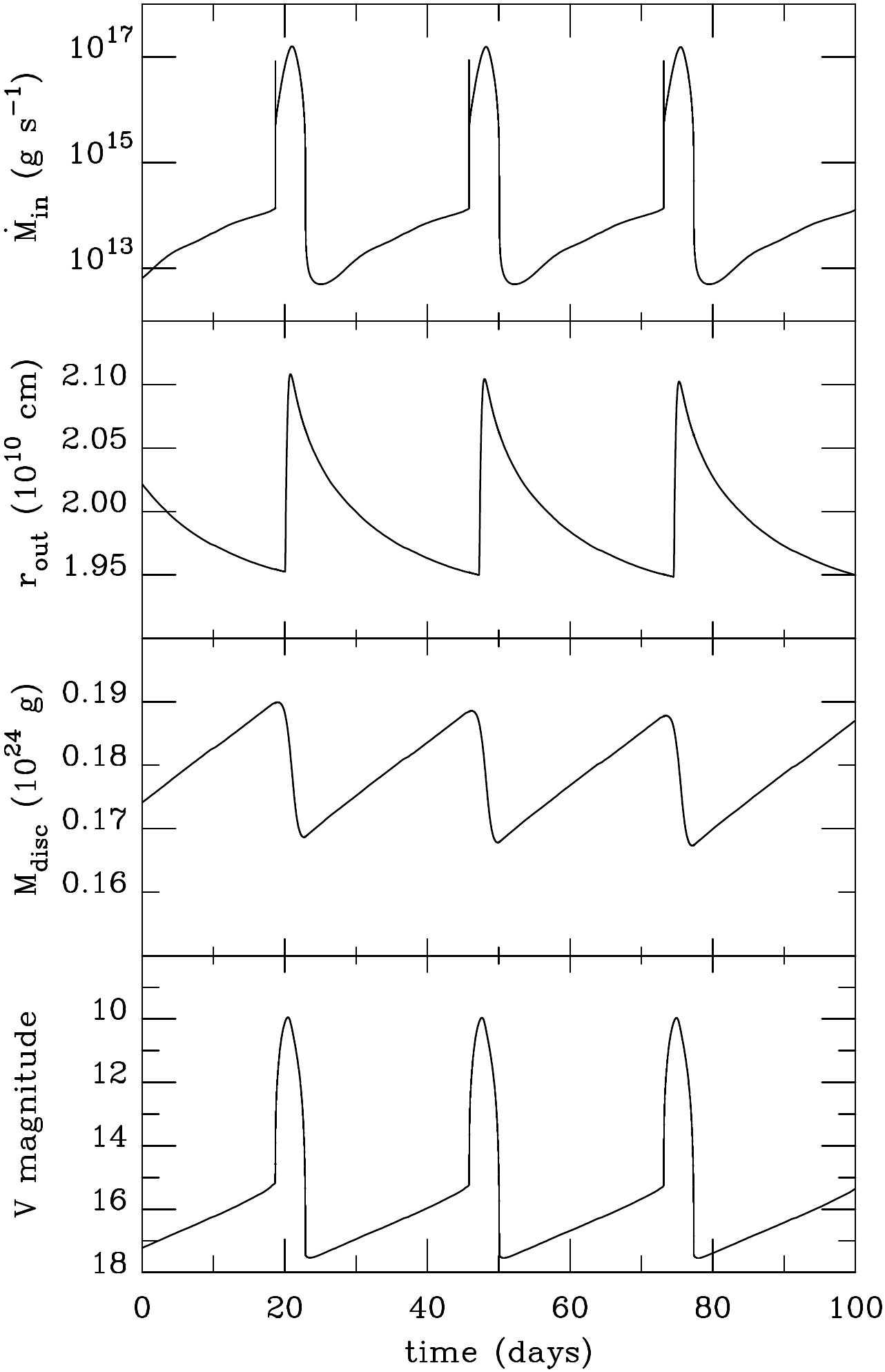}
\caption{Outbursts properties for $M_1 = 0.6$ M$_\odot$, $r_{\rm in} = 8.5 \times 10^8$ cm, $\alpha_{\rm cold} = 0.04$, $\alpha_{\rm hot} = 0.20$, $<r_{\rm out}> = 2 \times 10^{10}$ cm, and $\dot{M} = 10^{16}$ g s$^{-1}$. The upper panel shows the mass accretion rate onto the white dwarf, the second one the outer disc radius, the third one the disc mass, and the lower panel the visual magnitude. Note that only the disc contribution to the optical light is included. \citep[from][]{hmd98}}
\label{fig:lcurves}
\end{figure}

When the disc is unstable, the DIM accounts well for the observed characteristics (outburst amplitude and duration, recurrence time) of typical dwarf nova systems. \citet{l01} provides an extensive discussion of the development of the thermal instability that we briefly summarize here.

\begin{figure}
\includegraphics[width=\columnwidth]{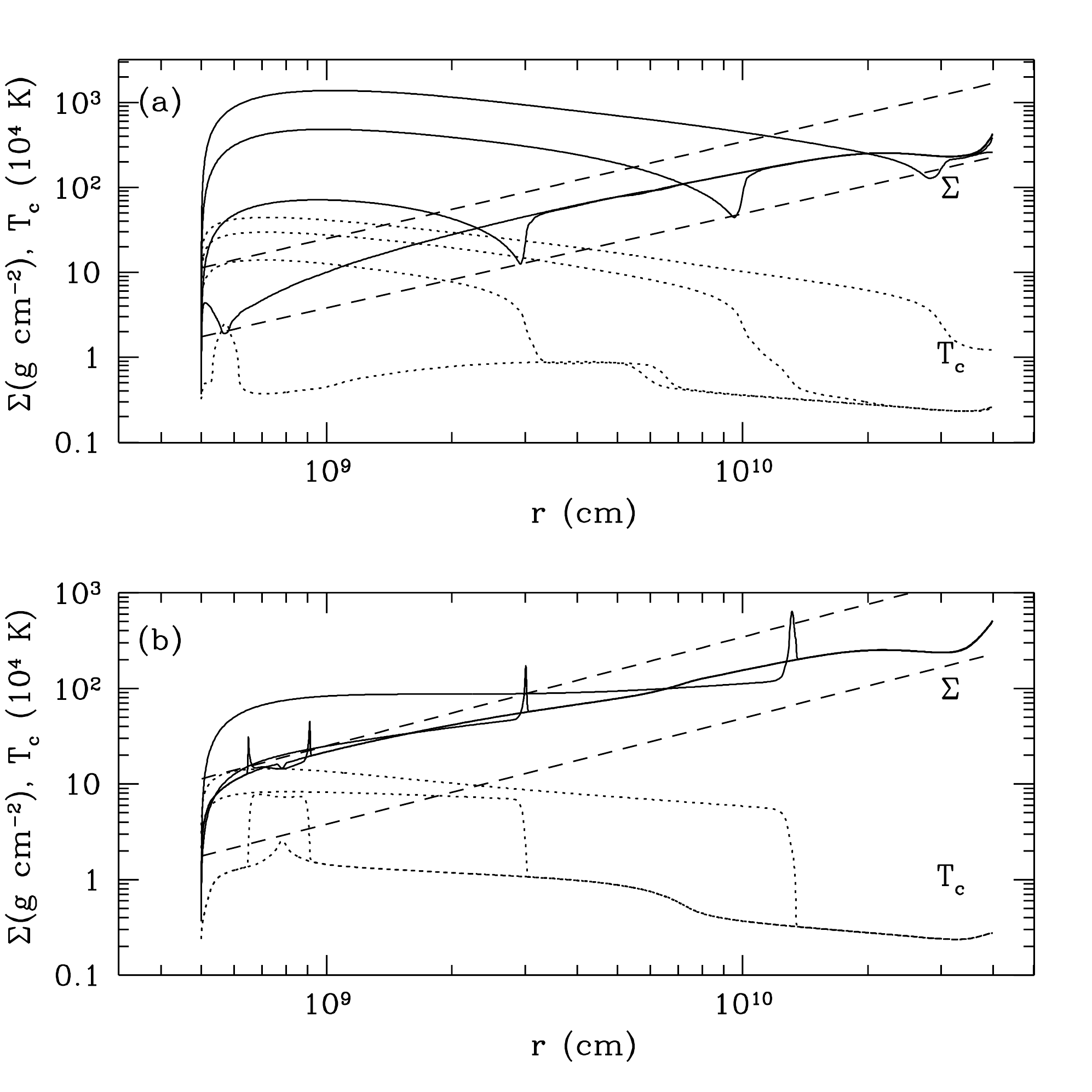}
\caption{Typical profiles of the surface density $\Sigma$ (solid lines) and the central temperature $T_{\rm c}$ (dotted lines) observed during the evolution of the thin disc. (a) Inward propagation of a cooling front. (b) Outward propagation of an inside-out heating front. Note the presence of a density rarefaction wave associated to a cooling front and a density spike in the case of a heating front. The dashed lines represent $\Sigma_{\rm min}$ (upper curve) and $\Sigma_{max}$ (lower curve). \citep[from][]{hmd98}}
\label{fig:profiles}
\end{figure}
The time evolution of the disc as predicted by the DIM is shown in Fig. \ref{fig:lcurves}.  

During quiescence, matter accumulates in the disc and slowly diffuses inwards; the disc mass builds up until, somewhere in the disc, the local mass accretion rate reaches $\dot{M}_{\rm crit}^+$. For high mass transfer rates, when the disc is close to be stable, this occurs at relatively large distances from the accreting object (but in any case, never close to the disc outer edge); the outburst is said to be outside-in; in most cases, this occurs very close to the disc inner edge, and the outburst is of the inside-out type. In both cases, two heat fronts propagate, one inwards that reaches rapidly the inner disc edge (after a few minutes), and one outwards that may reach the disc outer edge if there is enough mass in the disc, i.e. if the mass transfer rate is large enough. These fronts propagate at a speed of the order of $\alpha_{\rm h} c_{\rm s}$, where $c_{\rm s}$ is the sound speed estimated for a temperature close to that appearing in Eq. (\ref{eq:alpha}) (note that this is, as far as order of magnitudes are concerned, not very different from the sound speed in the hot medium), and their width is of the order of the disc scale height, which makes the thin disc approximation questionable \citep{mhs99}. 
The temperature raises, which decreases the viscous time, and at the peak of the outburst, the disc is not far from steady state; the surface density then scales approximately as $r^{-3/4}$ as expected from the Shakura-Sunyaev solution. During decline, $\Sigma$ decreases everywhere in the disc almost in a self-similar way, and eventually reaches $\Sigma_{\rm min}$ at the disc outer edge, since $\Sigma_{\rm min}$ varies as $r^{1.11}$. Then, a cooling front  starts from the outer edge and propagates towards the white dwarf, bringing the disc into quiescence. Cooling fronts propagate slower (by typically a factor of 10) and are broader than heating fronts. The outer disc radius changes by only 10\% durinng the outburst cycle, but this change has important consequences on the overall light curve of the system, as shown by \citet{hmd98}.

Figure \ref{fig:profiles} shows the time evolution of the surface density and temperature during an outburst. Here, the primary mass is 1.2 M$_\odot$, and the inner disc radius is 5000 km, slightly larger than the white dwarf radius (3900 km); the average mass transfer rate is $10^{-9}$ M$_\odot$ yr$^{-1}$, with a peak accretion rate of about $10^{-8}$ M$_\odot$ yr$^{-1}$ and a typical quiescence accretion rate of $2 \times 10^{-13}$ M$_\odot$ yr$^{-1}$. The average outer radius is $4 \times 10^{10}$ cm. The outburst starts at $r=8 \times 10^8$ cm. Note the steep density and temperature gradients, that require a good numerical resolution; thus equally spaced in $r^{1/2}$ grids that have been used in the past, because the disc equations take a simpler form when $r^{1/2}$ is used as a variable, are inadequate.

The observational difference between outside-in and inside-out outbursts is not obvious. It has long been thought that asymmetric profiles and lags between the UV and optical rise (the so-called UV delay) were the signature of outside-in outbursts; \citet{shl03} showed that the UV delay is slightly longer for inside-out than for outside-in outbursts, and that it is not a good indicator of the outburst type. In any case, the very notion of UV delay is ill-defined, because the rise both in the optical and in the UV is not instantaneous.

Except when the accretion rate is high and the outbursts are of the outside-in case, the recurrence time is equal to the viscous diffusion time, and thus independent of the accretion rate, but inversely proportional to $\alpha_{\rm c}$, thereby enabling in principle an estimate of $\alpha_{\rm c}$.

The DIM also predicts that the disc outer radius decreases during quiescence, because the disc is fed by matter with low angular momentum -- the circularization radius must be smaller than the outer disc radius. On the other hand, the disc expands during outbursts, because of the large transfer of angular momentum towards the outer edge resulting from the accretion of important quantities of matter at the inner edge. Analyses of the light curves of eclipsing systems that reveal the hot spot position and luminosity do show that the disc indeed expands during outbursts and contracts during the quiescence intervals\citep{s84,o86}. The finding by \citet{efc07} that, during a deep quiescence phase in U Gem, a symmetric full disc extends to $0.61 a$, close to the distance of $L_1$ from the center ($0.63a$) and larger than the average Roche radius of the secondary ($0.47a)$ does not fit with with this picture. One should nevertheless note that (i) the disc extension found by \citet{s01} in U Gem never exceeded $0.5a$, and (ii) that such a large disc would in any case pose severe problems from a dynamical point of view.

The DIM should also be tested against the predicted spectral evolution during a full cycle: quiescence, raise to outburst, maximum, decline and return to quiescence. The DIM provides the time evolution of the effective temperature at every point in the disc, and this can be used to produce theoretical time-dependent spectra that can in principle be compared to observations. For reasons that will be detailed in sect. \ref{sec:spectra}, and despite the existence of a wealth of observational data, not much has been done along this line after the initial work by \citet{ck87}, and has been rather concentrated on wavelength integrated properties, and more specifically on the UV -- optical timing properties as mentioned above.

Observations of cataclysmic variables using eclipse mapping techniques \citep{h85} provide the radius dependence of the effective temperature and hence the local accretion rate that can also be compared to theoretical predictions. The results \citep[see e.g.][]{b16} do not agree well with the DIM, as the temperatures are often found to be too high in quiescence, with, in the case of EX Dra, a radial distribution corresponding to steady state. As noted by \citet{h85} and emphasized later by \citet{s94}, for high inclination systems -- the inclination of EX Dra is 85°, the assumption of a flat disc leads to distortions of the reconstructed temperature-radius relation. In the case of Z Cha, that has a smaller inclination (81°), \citet{hc85} found that the disc was not far from being steady during an outburst.

\section{Extensions of the model}
The model does not, however, reproduce the long recurrence times of soft X-ray transients; more generally, when the accretor mass is large or when the inner radius is small, an alternation of many small and large outbursts is expected, which is not observed, and the quiescent luminosity becomes extremely small. To avoid these undesirable effects, the DIM must be slightly modified.

\subsection{Truncation of the inner disc} \label{sec:trunc}
The inner accretion disc will be truncated if the white dwarf possesses a magnetic field strong enough to disrupt the accretion flow close to its surface.  This occurs at the Alfvén radius, given by \citep{do73}:
\begin{equation}
r_{\rm mag} = 2.66 \times 10^{10} \mu_{33}^{4/7} M_1^{-1/7} \left( \frac{\dot{M}_{\rm acc}}{10^{16} \; \rm g s^{-1}} \right)^{-2/7} \; \rm cm,
\end{equation}
where $\dot{M}_{\rm acc}$ is the accretion rate onto the white dwarf and $\mu_{33}$ is the white dwarf magnetic moment in units of $10^{33}$ Gcm$^3$. 

The disc can also be truncated even in the absence of a magnetic field; this is required by observations of black hole soft X-ray transients in quiescence, as e.g. V404 Cyg for which \citet{brs16} find that the inner disc radius is larger than $3.4 \times 10^4$ times the gravitational radius in quiescence. Such a truncation can be due, in the black hole case, to the formation of a hot, optically thin, radiatively inefficient flow \citep[see e.g.][]{lny96,nbm97,nm08} and/or to the formation of a jet. In this context, truncation therefore means transition from a classical, geometrically thin and optically thick flow to a hot, optically thin flow. The precise nature of this hot inner flow has no importance as far as the disc stability properties are concerned, as long as it does not interfere with the outer classical thin disc.

Similarly, observations imply \citep{balman15} and models predict  \citep[see e.g][]{lmm97} the formation of a central hole in the accretion disc of dwarf novae during quiescence. In 8 systems, \citet{balman15} found large truncation radii during quiescence, of order of $(3 - 10) \times 10^9$ cm, quite larger than the truncation radius sometimes found in bright systems \citep{bgs14,gsb17}, of the order of $1.2 - 2$ times the white dwarf radius, for which an alternative option could be the existence of magnetically controlled coronal zones \citep{np19}.

One important effect of disc truncation is the suppression of short and weak outbursts that are produced by the DIM when the mass of the accretor is large. These outbursts are too weak to empty the disc -- the whole disc is not brought into the hot state, and they occur between major outbursts. Another important effect is the possibility that some systems lay on the cold, stable branch. \citet{hl17} investigated the predictions of the DIM in the case of intermediate polars (IPs) and found that the presence of a magnetic field can indeed account for the deficit of outbursting intermediate polars (see also Fig. \ref{fig:mdot_obs}). They also showed that the DIM cannot account for the infrequent and short (less than one day) outbursts observed in systems such as TV Col; these must be produced by some other mechanism, possibly an instability coupling the white dwarf magnetic field with the field generated by the MRI in the disc.

\citet{mhl00} showed that, in the case of soft X-ray transients, the truncation of the inner disc is not sufficient in itself to account for the observed properties of these systems; an additional ingredient must be added, disc irradiation, as we shall see later in sect. \ref{sec:disc_irr}.

\subsection{Mass transfer variations}
\subsubsection{Intrinsic mass transfer variations}
The mass transfer rate from the secondary $\dot{M}_{\rm tr}$ is observed to vary in AM Her systems, that do not have an accretion disc and in which the luminosity variations directly reflect variations of $\dot{M}_{\rm tr}$. There is no reason that such variations would not exist in other systems, and it had been proposed by \citet{o70} and by \citet{bep74} that dwarf nova outbursts were mass transfer bursts due to an instability in the outer envelope of the secondary star. Although these instabilities are no longer considered as an explanation for the dwarf nova phenomenon, significant mass transfer fluctuations must exist, although not much is known about their origin. 

Three types of mass transfer variability intrinsic to the secondary have been put forward: low amplitude variations, mass transfer outbursts and sudden drops of $\dot{M}_{\rm tr}$, similar to those observed in AM Her systems. \citet{lp94} proposed that star spots passing in the $L_1$ region can efficiently block mass transfer and account for the low states observed in AM Her systems as well as in other systems, such as the VY Scl stars. Not much is known about the mass transfer outbursts; they could be responsible for the stunted outbursts discussed later (see sect. \ref{sec:stunted}), or could trigger some dwarf nova outbursts, as proposed by \cite{lhh95}.

Moderate mass transfer variations occur in Z Cam systems; the standstills corresponds to periods when $\dot{M}_{\rm tr} > \dot{M}_{\rm crit}^+$, whereas normal outbursts are obtained when the transfer rate is lower. In these systems, $\dot{M}_{\rm tr}$ is close to $\dot{M}_{\rm crit}^+$, so that small variations by 10 -- 30 \% are sufficient to account for the Z Cam phenomenon; the origin of such small variations is unknown, and possibly of multiple origin. A firm prediction of the DIM is that standstills cannot be terminated by an outburst, because the disc is already in a high state, but must instead end by a return to quiescence. This is indeed what happens in most case

\subsubsection{Irradiation-induced mass transfer variations}
Mass transfer variations may also be extrinsic to the secondary star, and may be, for example, due to irradiation of the secondary by the variable accretion luminosity. \citet{o85} proposed that the superoutbursts in SU UMa systems are due to an irradiation induced mass transfer instability before proposing the tidal-thermal instability model that will be discussed later (see sect. \ref{sec:tti}). Variations of the mass transfer rate as a result of the secondary irradiation have been invoked in a large variety of contexts, ranging from supersoft sources (SSS) and T Pyx \citep{kkp00}, soft X-ray transients \citep{hkl86b}, WZ Sge systems \citep{hlh97}, VY Scl systems \citep{www95}, etc. In the specific case of dwarf novae, mass transfer from the secondary is observed to increase by factors up to 2 \citep{s95} during outbursts, and it is tempting to relate these mass transfer variation to irradiation of the $L_1$ region. The outcome of such irradiation is rather controversial; \citet{s04}, \citet{vh07} and \citet{c15} came to different conclusions concerning the magnitude of the effect. Parametrizing our ignorance by setting the mass transfer rate from the secondary $\dot{M}_{\rm tr} = \max (\dot{M}_{\rm tr,0},\gamma \dot{M}_{\rm acc})$, where $\dot{M}_{\rm tr,0}$ is the mass transfer rate from the secondary in the absence of irradiation and $\gamma$ a constant that must be smaller than unity, \citet{hlw00} were able to reproduce a large variety of light curves, including the alternation of normal outbursts and superoutbursts.

\subsection{Disappearance of the disc}
If the magnetospheric radius becomes larger than the circularization radius $r_{\rm k}$, the disc can no longer exist in a steady state since the boundary condition (\ref{eq:bc2}) cannot be satisfied. If one starts from an initially well established accretion disc with a small magnetospherical radius and then slowly increase $r_{\rm mag}$, the outer radius decreases, and eventually reaches $r_{\rm k}$. This happens precisely when $r_{\rm mag} = r_{\rm k}$; at this point, the excess of angular momentum to be removed by tidal torques vanishes. In the case of a magnetized accreting white dwarf, this occurs if the mass transfer rate is less than a critical value:
\begin{equation}
\dot{M}_{\rm tr} < 4.89 \times 10^{16} M_1^{-0.72} \mu_{33}^{0.87} \; \rm g s^{-1}.
\label{eq:mdotmax}
\end{equation}
There are observational evidences that many intermediate polars do possess an accretion disc; it was however proposed by \citet{hkl86} that in some steady IPs, the magnetic field may prevent the formation of a disc \citep[see also][]{kl91}. Condition (\ref{eq:mdotmax}) can also be fulfilled during low states. \citet{hl17b} proposed that the absence of outbursts during the decline of FO Aqr from the normal state, during which the accretion disc is observed to be present and the accretion rate is high enough for the system to be stable on the hot branch, to a low state can be explained only if the accretion disc disappears before the system enters the unstable region. If this were not the case, and because the mass transfer rate in FO Aqr changes on time scales that are long as compared to the viscous time, the system should be in the dwarf nova instability strip during a fraction of the long decline time to the low state. \citet{hl17b} showed that this must happen when the system visual magnitude is of order of 14. Interestingly, \citet{lgk19} found that a transition in the accretion mode occurred when the luminosity dropped below $V=14$, at which point there was a strong and persistent interaction between the accretion stream and the white dwarf magnetosphere.

A similar explanation was put forward by \citet{hl02} to account for a similar absence of outbursts in the low states of VY Scl systems. This explanation requires that these systems harbor a magnetic white dwarf. As noted by \citet{zom14}, this explanation would account for the lack of a clear correlation between the UV/optical and the X-ray observations because the X-rays would be emitted by the accretion flow on the polar caps, while the optical would originate from the accretion disc. \citet{smh18} found a ring like structure in the H$\alpha$ and H$\beta$ Doppler maps of VY Scl during a low state, and concluded that the very presence of this structure proves the existence of an accretion disc. Whereas this observation clearly proves the existence of material orbiting the white dwarf, the velocities ($\leq 400$ km/s) correspond to distances of order of $6.4 \times 10^{10}$ cm for a 0.8 M$\odot$ white dwarf. This is slightly larger than the tidal truncation radius, $4 \times 10^{10}$ cm, meaning that the origin of the line is unclear. One should note that one expects that some material may orbit the white dwarf and form a narrow ring even if the stream directly impacts the magnetosphere, but the ring should be located at distances close to the circularization radius, of order of $1.2 \times 10^{10}$ cm, corresponding to much higher velocities than observed.

\begin{figure}
\includegraphics[width=\columnwidth]{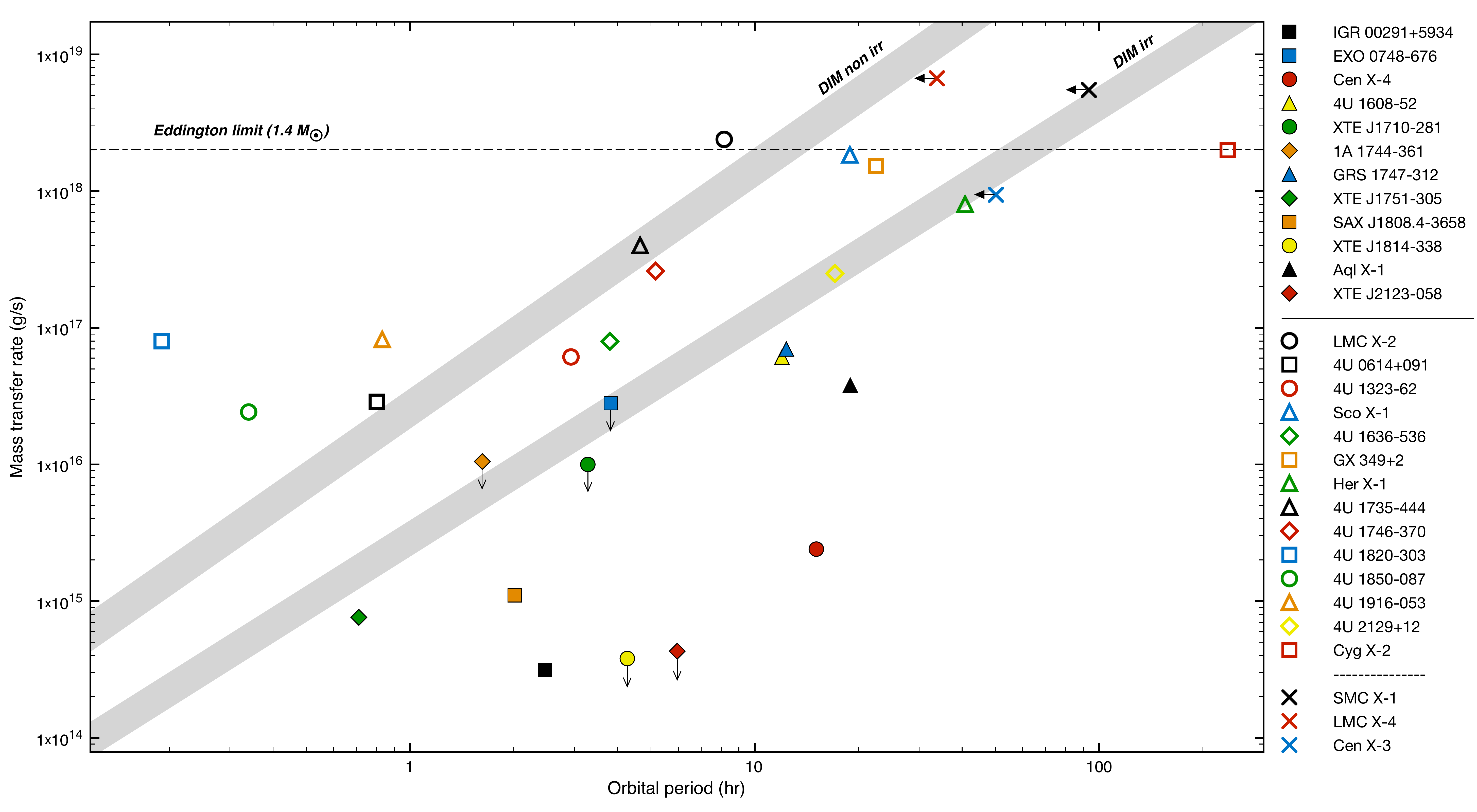}
\caption{Mass transfer rate as a function of the orbital period for XRBs with neutron stars. The transient and persistent LMXBs have been indicated with filled and open symbols respectively, while the crosses indicate the high-mass, persistent systems. The shaded grey areas indicated `DIM irr' and `DIM non irr' represent the separation between persistent (above) and transient systems (below) according to the disc instability model when, respectively, irradiation is taken into account and when it is neglected. The horizontal dashed line indicates the Eddington accretion rate for a 1.4 M$_\odot$ neutron star. Similar results are obtained when the accretor is a black hole \citep[from][]{cfd12}.}
\label{fig:sxt}
\end{figure}

\subsection{Irradiation of the accretion disc} \label{sec:disc_irr}
\subsubsection{Irradiation by the accretion luminosity}
In X-ray binaries, the accretion disc is observed to be significantly affected by irradiation, and it had been early recognized \citep{vm94,v96,kr98,dlh99} that X-ray irradiation of the disc must significantly change the stability properties of the accretion disc. The effective temperature of the disc is increased, and, if brought above a value of order of typically 8,000 K, the cold and intermediate branches of the S-curve can no longer exist, since hydrogen is fully ionized. 

In principle, the vertical structure of the disc should be determined in a self-consistent way, including in particular the screening from the central source of the outer regions by the innermost, heated and therefore inflated portions of the disc. Confirming \citet{tmw90}, \citet{dlh99} found that self-screening prevents the outer parts of the disc to see the central source, contrary to observations. The disc must therefore be warped, or the central irradiating source must be extended, or both. The source extension can be either due to the formation of a hot corona or to the existence of thermal winds as considered by \citet{ddt19} that scatter the X-rays back to the disk. The problem then becomes extremely complex, and depends on many unknown parameters such as the shape of the disk or the source extent; \citet{dlh99} used instead the prescription 
\begin{equation}
T_{\rm irr}^4 = {\cal C} \frac{\eta \dot{M_{\rm acc} c^2}}{4 \pi \sigma R^2}
\end{equation}
where $\eta \sim 0.1$ is the accretion efficiency and $\sigma$ the Stefan-Boltzman constant. \citet{dhl01} found that using a constant $\cal C$ of order of $5 \times 10^{-3}$ would reproduce the properties of soft X-ray transients.
When irradiation is taken into account, the critical accretion rate becomes  \citep*[][appendix A]{ldk08}:
\begin{equation}
\label{mcrit-irr}
\begin{aligned}
\dot{M}_{\rm crit}^+ & = 9.5 \times 10^{14} 
                    ~{\cal C}_{-2}^{-0.36}
                    ~\alpha_{0.1}^{ 0.04+ 0.01\log{\cal C}_{-3}}
                    ~R_{10}^{2.39-0.10\log{\cal C}_{-2}} \\
                    &  \times   ~M_1^{-0.64+ 0.08\log{\cal C}_{-2}}\,\rm g\,s^{-1}
\end{aligned}
\end{equation}
where $C = 10^{-2} \,  C_{\rm -2}$.
\citet{cfd12} found that, when irradiation is taken into account, the critical $\dot{M}$ separating transient and steady X-ray binaries agrees well with the model, as shown by Fig. \ref{fig:sxt}.

For soft X-ray transients, irradiation of the outer disc is not sufficient for reproducing the light curves. \citet{dhl01} showed that, unless the disc does not extend down to the neutron star surface or to the innermost stable orbit in the case of a black hole, but is instead truncated as discussed in sect. \ref{sec:trunc}, multiple reflares are unavoidable. On the other hand, if both truncation and irradiation are taken into account, the light curves produced by the model are in good agreement with observations.

In the case of accreting white dwarfs, one can use the same prescription as for the X-ray binaries when heating is due to EUV emission from the boundary layer. However, because the potential well of a white dwarf is much shallower than that of a neutron star or a black hole, \citet{bhl18} showed that the irradiation flux is much smaller than the viscous flux, even at large distances, unless the white dwarf is close to the Chandrasekhar limit.

\subsubsection{Irradiation by a hot white dwarf}
If the white dwarf is hot, there is an additional heating term \citep[see e.g.][]{hld99}:
\begin{equation}
T_{\rm irr}^4 = (1- \beta)\left[\sin^{-1}\rho - \rho\sqrt{1 - \rho^2}\right]\frac{T^4_{\rm WD}}{\pi},
\label{eq:irrwd}
\end{equation}
where $\beta$ is the disc albedo, $\rho=R/R_{\rm WD}$, and $R_{\rm WD}$ and $T_{\rm WD}$ are the white dwarf radius and temperature respectively. Irradiation by the boundary layer is usually not very important, unless the primary mass is large, but irradiation by a hot white dwarf can be significant in particular during the low states \citep[see e.g.][]{lhk99,hl02}; during the high states, the accretion disc luminosity exceeds that of the white dwarf, and irradiation by the white dwarf itself is not significant.

The situation in symbiotics is different, although the physics is the same, because the accretion rate is much larger than in cataclysmic variables and can exceed the limit above which hydrogen burns in a stable manner; the white dwarf then becomes extremely hot. \citet{bhl18} considered the case of the symbiotic star Z And in which the white dwarf luminosity is of order of 10$^3$ L$_\odot$ and showed that irradiation strongly modifies the dwarf nova outburst properties, leading them to conclude that the observed outbursts were triggered by mass-transfer enhancements from the secondary star, leading to an increase of nuclear burning at the white dwarf surface.

\begin{figure}
\includegraphics[width=\columnwidth]{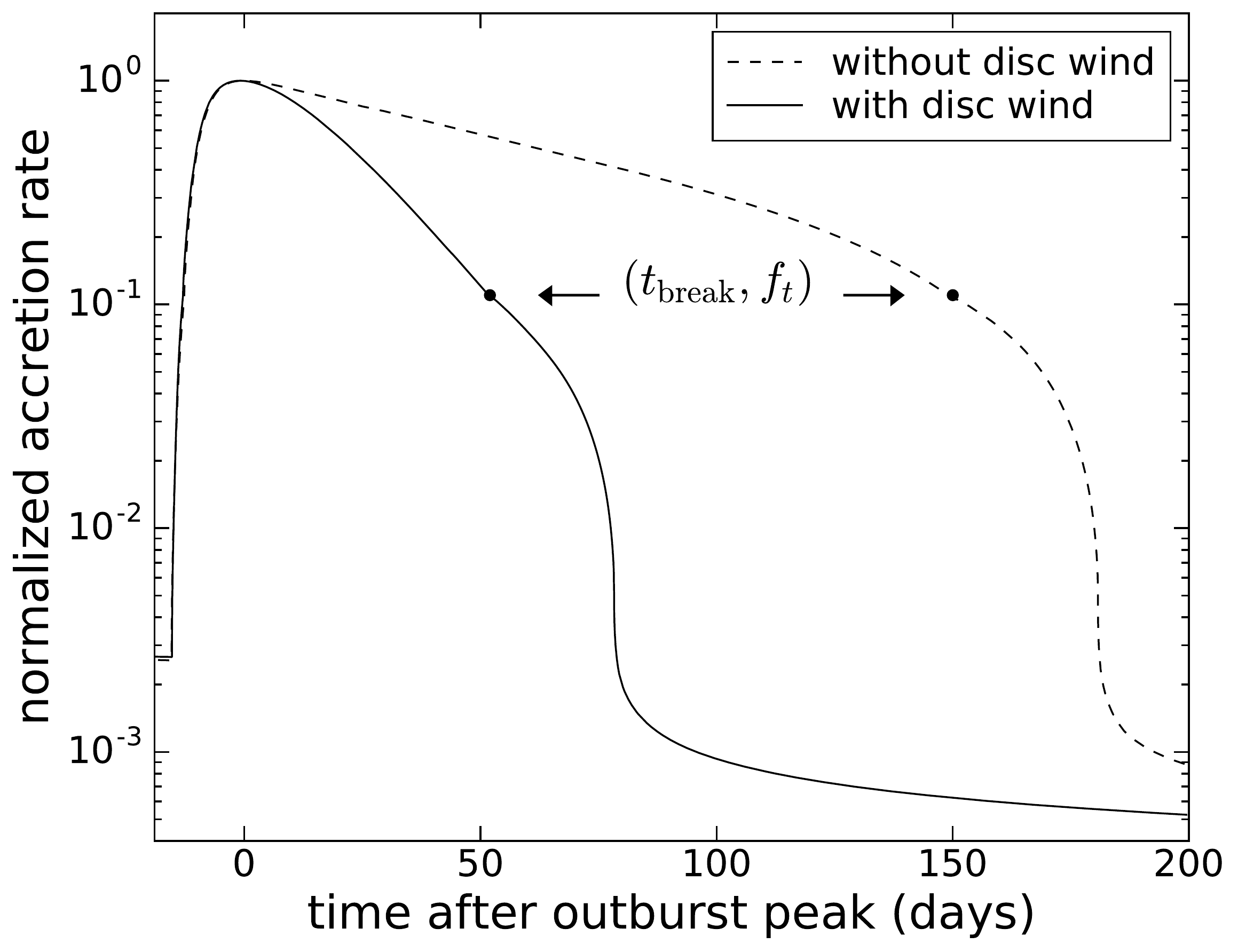}
\caption{Two model light-curves for an irradiated disc around a $6M_{\odot}$ black-hole and an $\alpha=0.2$ accretion disc are shown: (dashed line) no mass loss present, and (solid line) including a mass loss term during outburst. \citep[from][]{tlh18}. Reprinted by permission from Springer Nature.}
\label{fig:toy_model}
\end{figure}
\begin{figure}
\includegraphics[width=\columnwidth]{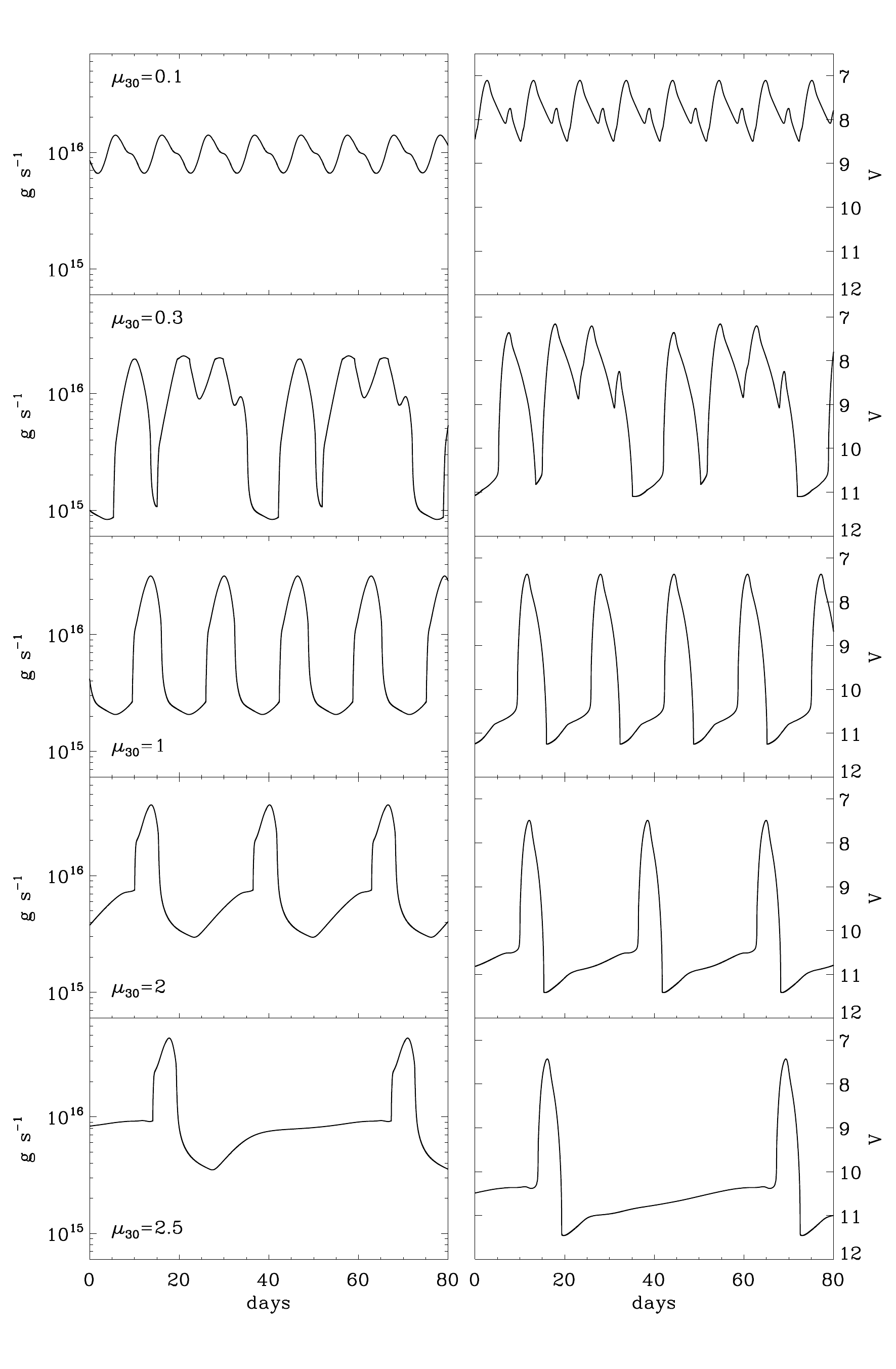}
\caption{Light curves obtained by \citet{sdl19} in the case where the vet vertical magnetic field threading the accretion disc has a dipolar variation; the left panel shows the mass accretion rate onto the white dwarf and the right panels show the absolute V magnitude. The magnetic moment is, from top to bottom, $\mu = [0.1, 0.3, 1, 2, 5] \times 10^{30}$ G cm$^3$. Reproduced with permission from Astronomy and Astrophysics \copyright ESO.}
\label{fig:dipole_lc}
\end{figure}

\subsection{Winds and outflows} \label{sec:winds}
Winds and outflows are observed in many if not all accreting systems, including cataclysmic variables \citep[see][and references therein]{mkl15} and X-ray binaries \citep[see][for a review]{db16}. These winds carry both mass and angular momentum, which should significantly affect the thermal/viscous instability. The total amount of mass-loss in the wind can be a significant fraction of the mass transfer rate; this is certainly the case when the luminosity approaches or even exceeds the Eddington limit in soft X-ray transients or in symbiotic binaries, but one should note that a high luminosity is not a necessary condition for the formation of a strong wind. \citet{tlh18} assumed that the local wind mass loss rate from the disc is proportional to the local accretion rate and that matter is lost with its specific angular momentum. The resulting light curve they obtained using the DIM is shown in Fig. \ref{fig:toy_model} in the case of a 6 M$_\odot$ black hole. The shape remains the same as in the no-wind case, but the decay time-scale is significantly reduced. This is equivalent to increasing the viscosity parameter $\alpha$ in the hot state to values significantly larger than the standard 0.1 -- 0.2 deduced from observations in cataclysmic variables; they argue that this effect might be responsible for the high values of $\alpha$ they obtain when fitting the light curves of 12 transient systems that give $\alpha$ in the range 0.19 -- 0.99. Of course, as they note, there need not be a correlation between the local mass loss rate and mass accretion rate, and a more detailed modelling is obviously required. Note that the origin of the wind is not specified; its presence during the whole outburst duration suggests a magnetic origin \citep{tlh18}.

In contrast to the previous model, \citet{sld18b} considered the case where the wind is magnetized and exerts a torque on the accretion disc. They also assume that mass loss via the wind is small as compared to the accretion rate, which is a good assumption for cataclysmic variables. Moreover, because a net vertical magnetic flux is assumed, the MRI is enhanced, which leads to an even more efficient angular momentum transport. While the MRI can be reduced to an $\alpha$ prescription, this is not the case for  the wind torque, because angular momentum transport is not accompanied by heating. They found that the MRI can be quenched by resistivity on the cold branch, unless the magnetic field is strong, but in this case the effective $\alpha$ should then have the same value as on the hot branch, typically $\alpha = 0.1$. They argued that a more likely outcome is that in quiescence, angular momentum transport is dominated by wind-driven angular momentum transport. \citet{sdl19} then calculated the expected light curves predicted by the DIM plus a wind-driven angular momentum transport, using a temperature-dependent $\alpha$ that fits their MRI results, and assuming a vertical magnetic field that can be either independent on radius or dipolar. They find (see Fig. \ref{fig:dipole_lc}) light curves that are not dissimilar from those of dwarf novae if the field has a dipolar dependence, whereas a constant magnetic field would not lead to outbursts but merely to small oscillations of the mass transfer rate and visual magnitude. Note that, although dipolar, the magnetic field in the disc is not connected to the white dwarf (i.e. no propeller effect is included). 

These two examples show that winds might be extremely important, and should be included in simulations. But the net magnetic field should be calculated self consistently, from both the coupling of the white dwarf magnetic field, and the transport magnetic field from the secondary by the accretion flow. However, since the primary and secondary magnetic fields show large variations from systems to systems, one would also imagine that the stability properties of cataclysmic would depend on the mass transfer rate, plus another parameter; this does not seem to be the case, as shown by Fig. \ref{fig:mdot_obs}. 

\subsection{Helium secondaries}
AM CVn are binary systems in which a white dwarf accretes matter from a helium secondary. These systems may also have outbursts, similar to dwarf novae. \citet{s83} calculated the vertical equilibrium structure of helium discs, and showed that, as for hydrogen discs, S curves are found, so that the DIM should also apply to these systems. The main difference with hydrogen discs is that, because of the higher ionization temperature of helium, the turning points of the S curve correspond to higher effective temperatures, 9700 K and 13,000 K instead of 5200 and 6900 K. Soon after, \citet{c84} and later on \citet{to97} developed models that showed that the DIM could indeed account for the observed bursts once the changes in chemical composition had been implemented; a more refined version of the model was then put forward by \citet{kld12} who showed that the same modifications to the DIM that had been introduced in the DIM for CVs (irradiation, disc truncation) had also to be included in this case; in particular,  the  enhanced mass-transfer rate,  due presumably to variable irradiation of the  secondary, must not only be taken into account but is a factor that  determines the shape of most AM CVn outburst light-curves.

\subsection{Tidal-thermal instability}
\label{sec:tti}
SU UMa systems are dwarf novae that show, in addition to normal outbursts, long and bright outbursts called superoutbursts that occur regularly, typically every few normal outbursts, and last about ten times the duration of normal outbursts. In addition to the superoutburst phenomenon, they also show superhumps during superoutbursts, that are modulation of the light curve at a period longer by a few percent than the orbital period.

\citet{o89} proposed that superoutbursts in SU UMa systems are due to the fact that, when the accretion disc outer edge reaches the 3:1 resonance radius, a tidal instability develops, that results in an eccentric precessing accretion disc, with enhanced angular momentum transport. The disc precession would be responsible for the superhumps observed in the light curve during superoutbursts, while the enhanced angular momentum transport would be the explanation for the longer duration and increased brightness of the superoutbursts as compared to normal outburst. In this model, a sequence of normal outbursts occurs; the disc mass slowly builds up and the disc radius increases until, during an outbursts, the disc radius reaches the 3:1 resonance radius at which point the tidal instability develops. The tidal torque suddenly increases and remains large until the disc has shrunk to a value that is smaller than the 3:1 resonance radius by typically a factor 1.5. As noted earlier, the disc outer radius grows during normal outbursts, and decreases during the quiescent phases. The superoutbursts must therefore be triggered by a normal outburst. This model is thus a combination of the standard thermal-viscous instability and of a tidal instability, hence its name, the tidal-thermal instability model (TTI). It is supported by the evidence that the SU UMa systems are, with a couple of exceptions (TU Men and OGLE GD-DN-9), all below the period gap, and therefore have low mass secondaries, so that the mass ratio $q=M2_/M_1$ can be well below 0.3. The 3:1 resonance radius then sits within the Roche lobe of the primary, and the disc can indeed extend to the 3:1 resonance radius. The TTI model is also supported by 2D SPH simulations \citep{w88} that showed that for a system with a mass-ratio $q = 0.15$, the disc is tidally unstable and can become asymmetric and slowly rotate in the inertial frame of reference. The tidal stresses raised in the disc by the secondary produce periodic peaks in the light curve that are interpreted as superhumps \citep{ho90}. \citet{wk91} showed that other resonances (e.g. 4:1) that can occur in systems with not so extreme mass ratios do not lead to such behavior. These early simulations have been confirmed by later ones \citep[see e.g.][]{kpo08}.

I shall return in mode details to the applicability of the TTI model to SU UMa in Sect. \ref{sec:suuma}.

\section{A few open questions, unsolved problems and limitations of the DIM}
A number of difficulties discussed below are not specific to the DIM but much more general. However, the DIM inherits the deficiencies of the disc models, and these are sometimes exacerbated in the case of dwarf novae, as is for example the case for viscosity in the low state.

\subsection{Angular momentum transport} \label{sec:viscosity}
\subsubsection{Magnetorotational instability}
The first major problem that is not specific to the DIM is the nature of angular momentum transport. Values of $\alpha$ requested for the DIM to reproduce the characteristic outburst time-scales have been somewhat difficult to reconcile with those inferred from the results of numerical simulations of the MRI. Local shearing box simulations developed over the past 25 years have yielded values of $\alpha$ of the order of 0.03 \citep[see e.g.][]{sba12}, and higher values were obtained only when a net vertical magnetic flux was introduced \citep{hgb95}. 

\begin{figure}
\includegraphics[width=\columnwidth]{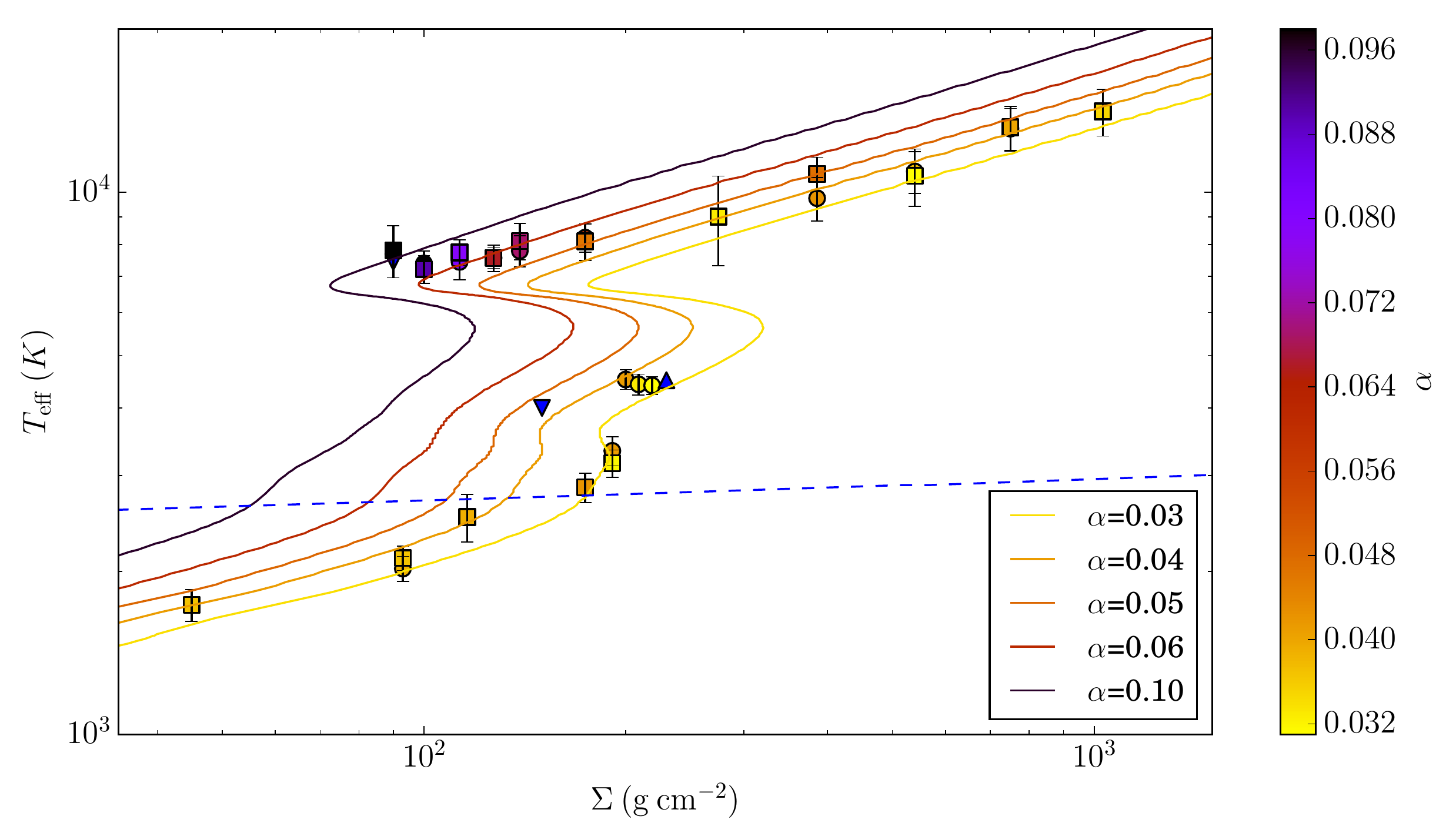}
\caption{Example of $\Sigma - T_{\rm eff}$ S-curves calculated from MRI simulations. The solid lines correspond to the standard $\alpha$ model; the dots indicated the results of MRI simulations and the color coding indicates the equivalent $\alpha$ \citep[from][]{sld18}. Reproduced with permission from Astronomy and Astrophysics \copyright ESO.}
\label{fig:scurve_mri}
\end{figure}

It was later realized that convection significantly affects the MRI, and that values of $\alpha$ of order of 0.1 could be obtained \citep{hbk14,cyb17,sld18}, but these are obtained only close to the upper turning point of the S curve; $\alpha$ returned to low values for higher temperatures on the upper branch because convection is no longer present. They did find however that $\alpha$ is not increased on the intermediate, unstable branch, although convection is present. In all cases, an S-shaped $\Sigma - T_{\rm eff}$ curve is obtained (see  Fig. \ref{fig:scurve_mri}), that is key for the DIM to work.

\citet{ckb16} introduced in the DIM an $\alpha$ prescription reproducing the results of these numerical simulations which simply changed the temperature dependence of Eq. (\ref{eq:alpha}), and they did find that these MRI-based models can successfully reproduce observed outburst and quiescence durations, as well as outburst amplitudes, albeit with different parameters from the standard disc instability models. However, details of the light curves did not agree with observations; in particular, reflares during the outburst decline were found in all cases, generating rebrigntenings of the light curve that are not observed, except in some very particular systems such as WZ Sge, and only at the end of a an outburst (see Sect. \ref{sec:rebr}).

Local shearing box simulations have their limitations though; it is in particular unclear if they are representative of the global dynamics. These simulations have periodic and semi-periodic boundary conditions that are clearly not justified in the radial direction (and in the vertical direction in the case of unstratified simulations). The impact of these assumptions, that change the mathematical nature of the problem needs to be further investigated. \citet{bp99} showed that local shearing box simulations inevitably gives rise to an $\alpha$ disc.

One can therefore conclude that the $\alpha$ prescription, although being a simplistic description of angular momentum transport in accretion discs, is not completely at odds with MHD simulations that do indeed provide a much better description of the flow dynamics, but cannot at present incorporate the thermodynamics with a sufficient accuracy to be able to account for the thermal instability.

One should however keep in mind that simulations have been most often performed assuming that non-ideal effects (resistivity and Hall effect) can be neglected. While this is certainly true on the hot branch, this is not the case on the cool branch, which lead \citet{gm98} to speculate that the MRI could be ineffective on the cold branch, thereby explaining the drastic change in viscosity occurring when the hydrogen becomes neutral. These non-ideal effects have been studied in the case of protoplanetary discs \citep[see e.g.][]{lkf14,b15}, but not much has been done for discs in CVs although they are important during quiescence \citep{ckb16}.

\subsubsection{Spiral waves}
The possible ineffectiveness of the MRI in cold discs lead to investigate other options for angular momentum transport. Angular momentum can be transported by non-local processes, such as e.g. waves. Global MHD simulations have therefore been developed \citep[see][for the specific case of cataclysmic variables]{jsz16,jsz17}, but at the expense of an oversimplified treatment of the thermodynamics: the gas is usually assumed to be either locally isothermal or an adiabatic equation of state is taken. 

\citet{jsz17} showed that spiral shocks transport about the same amount of angular momentum as the MRI when the Mach number $\mathcal{M}$ is of order of 10 or lower and the seed magnetic field corresponds to a ratio of gas to magnetic pressure of 400. A linear analysis indicates that waves tend to be exponentially suppressed as $\mathcal{M}$ increases, but might be important when there are sharp radial gradients in the surface density \citep{xg18}. 

On the other hand, observations \citep[see e.g.][]{pgm19} indicate that spiral shocks are present and might play a role in the accretion process, but their interpretation is not straightforward.

It is therefore likely that spiral shocks do not play a significant role for angular momentum transport in accretion discs for CVs, in particular in the low state, since the Mach numbers are expected to be high. This preliminary conclusion might have to be revised in the future though. 

\subsubsection{Hydrodynamical instabilities}
Pure hydrodynamical effects have also been put forward to account for angular momentum transport \citep[see][for a recent and detailed review]{fl19}; although models have most often been developed in the context of protoplanetary discs, they should be applicable to accretion discs in CVs and LMXBs, in particular during the low states when the disc is cold and the resistivity is large. These models include for example the baroclinic instability \citep{kb03} and the vertical shear instability \citep{ngu13}. The values of $\alpha$ predicted by these models are often lower than those resulting fro MRI and are typically of order $10^{-4} - 10^{-2}$. The applicability to astrophysical discs of the results, obtained using idealized simulations, needs however to be confirmed. As an example, the baroclinic instability is severely suppressed in the presence of a magnetic field \citep{lk11}.

\subsubsection{Winds and outflows}
As discussed in Sect. \ref{sec:winds}, winds can also carry away angular momentum from the disc.

\subsection{The low state}
It has been noted since long that the disc instability model has difficulties in accounting for the quiescent state, both on theoretical and observational grounds. \citet{s00} noted that, while all models predict that the luminosity should gradually increase during quiescence until an outburst is triggered, this is not observed. \emph{Kepler} observations have shown that there are even cases where the luminosity decreases during quiescence \citep[this is the case for V1504 Cyg,][]{oop16}. 

From a theoretical point of view, the temperature is low enough that the ionization degree is most probably not sufficient for the MRI to be effective, as noted in Sect. \ref{sec:viscosity} just above. Other possibilities have been investigated, but as noted earlier, none is widely accepted.  

One should finally keep in mind that the optical depth of the quiescent disc is not large, and that its vertical structure is therefore poorly calculated in most models.

For all these reasons, the low state is the weakest point of the DIM.

\subsection{SU UMa systems: TTI or irradation of the secondary ?} \label{sec:suuma}
Superoutbursts and superhumps have also been found in systems with long orbital periods, including U Gem \citep{sw04} that has a well determined mass ratio $q=0.36 \pm 0.02$ \citep{s01}, too large for the 3:1 resonance to be able to develop. \citet{c12} showed that in these long period systems, a superoutburst is also triggered by a precursor, as in normal SU UMa systems, making the hypothesis that there could be two different classes of superoutbursts, and hence two different explanations, difficult to hold. Moreover, questions were raised on the ability of the tidal instability model to account for the amplitude of the superhumps; \citet{s09b} argued that the predicted superhump amplitudes are ten times smaller than the observed ones. 

\begin{figure}
\includegraphics[width=\columnwidth]{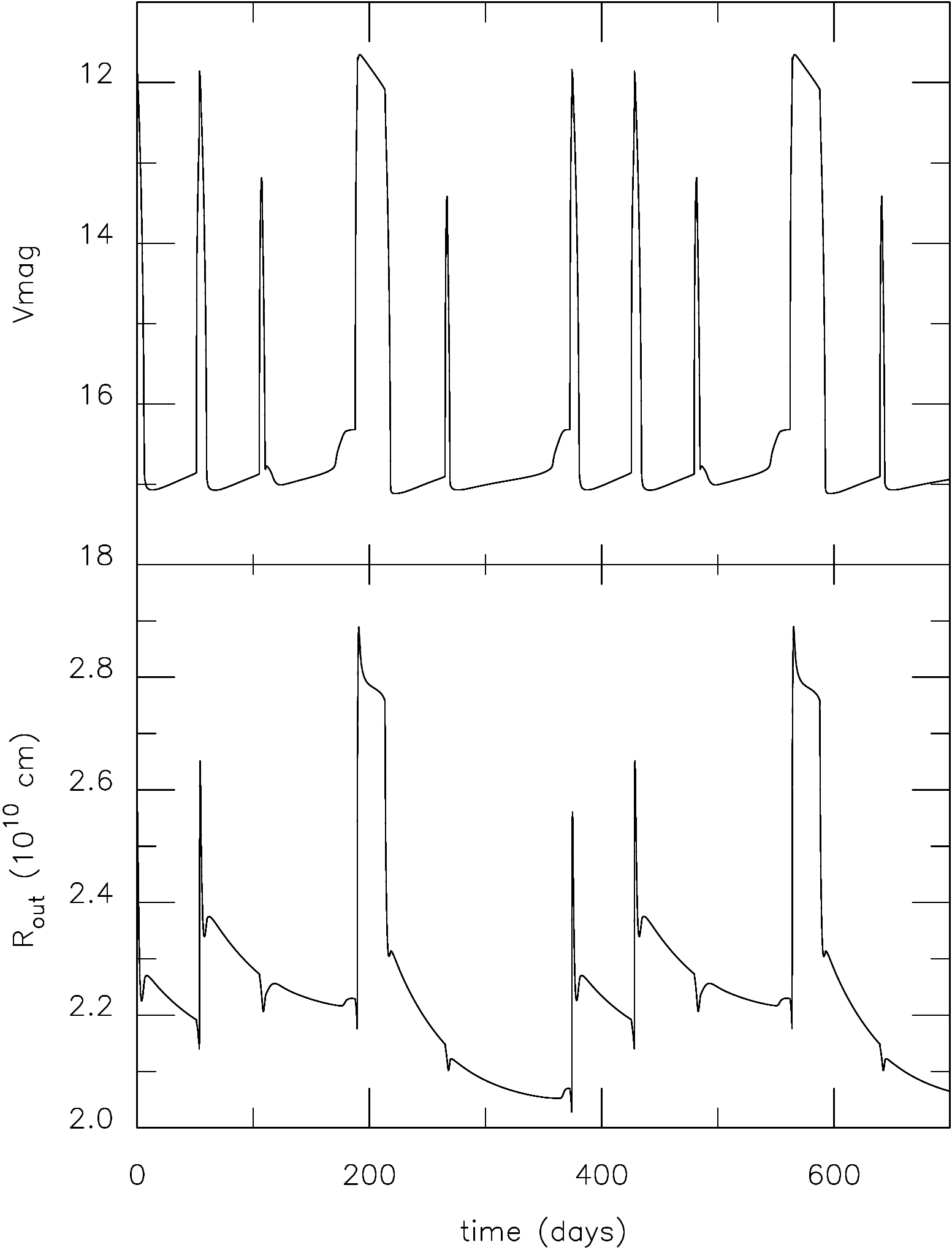}
\caption{Time evolution of the outer disc radius when the mass transfer rate is modulated by the secondary irradiation. Top panel: visual magnitude; bottom panel: outer disc radius.}
\label{fig:su}
\end{figure}

These arguments lead \citet{s09} to propose that superhumps and superoutbursts are due to an irradiation-induced enhancement of mass-transfer. The increase of the hot spot luminosity during superoutbursts that has been observed in a number of systems clearly indicates that the mass transfer rate significantly increases, even when one accounts for the variations of the outer disk radius and hence of the hot spot position. \citet{hlw00} showed that the SU UMa behavior could be quite well reproduced by taking into account the modulation of mass transfer from the secondary as a result of irradiation.

\citet{csh10}  even showed that a sequence of long and short outbursts that have the same characteristics as the SU UMa star V344 Lyr can be generated by the DIM alone without considering a mass transfer modulation from the secondary; this required however to use a ratio $\alpha_{\rm h}/\alpha_{\rm c}$ 10 times larger than usually assumed, with $\alpha_{\rm h}= 0.1$, a standard value, and $\alpha_{\rm c}= 0.0025$. The inner disc had also to be truncated, with an inner radius of $2 \times 10^{9}$ cm, about twice the white dwarf radius.

In this context, \citet{s09} proposed that superhumps are due to a modulation of the mass transfer by irradiation of the secondary. This model instead relies on the fact that the secondary is observed to be irradiated periodically at the superhump period, which results in a mass transfer modulated at that period, and thus in a variable energy dissipation of the interaction of the stream leaving $L_1$ with the accretion disc. The clock mechanism is basically the same as in the tidal model: a strongly asymmetric accretion disc. The fact that the superhump amplitude and the irradiation amplitudes are observed to be correlated \citep{s17} supports this model. This model is also able to account for the fact that some systems with long orbital periods, including U Gem \citep{sw04} have been observed to show superhumps although they cannot be subject to the 3:1 resonance.

Despite these arguments, the tidal-thermal model for superoutbursts is widely accepted. An important difference that has been put forward between both models is the variation of the disc radius during outbursts, since the TTI model predicts a strong contraction of the disc during superoutbursts. It is, however, not always realized that the irradiation-induced mass transfer enhancement model also predicts a contraction of the disc. In order to show this, I run simulations whose results are shown in Fig. \ref{fig:su} that displays the time evolution of a system in which the mass transfer rate is modulated  by the secondary irradiation according to $\dot{M}_{\rm tr} = 10^{15} + 0.5 \dot{M}_{\rm acc}$ g s$^{-1}$. The orbital period is 1.8 hr, the primary and secondary masses are 0.8 and 0.15 M$_\odot$ respectively; $\alpha$ is 0.1 on the hot branch and 0.02 on the cold branch. A sequence of long outbursts separated by four short outbursts is produced; for two of these outbursts, the heating front does not reach the outer edge of the disc and there is no radius expansion. As a result of the increased mass transfer rate, the disc contracts during outbursts after a possible rapid expansion phase. Note also that the large outbursts do not start with a precursor; the details of the light curve and of the radius variations depend on how the mass transfer rate is modulated by irradiation, which is poorly constrained \citep[such a precursor was found in][for slightly different parameters]{hlw00}.

Observations do show that the disc contracts during a superoutburst \citep{ok14}, but this contraction is not violent, and occurs only a few days after the onset of the superoutburst.

The observation of a superhump during a normal outburst of SU UMa \citep{iai12} also raises questions about the relation of superhumps and superoutbursts. This normal outburst occurred in the middle of a supercycle, a marked difference from V1504 Cyg that also showed superhumps during a normal outburst \citep{kmm12}, but this normal outburst occurred just before a superoutburst.

One must therefore seriously reconsider the validity of the TTI model, and remember that \citet{vp83} proposed a long time ago that the superoutbursts of the SU UMa stars are simply an extension of the long outbursts of normal dwarf novae at short orbital period. In fact, the DIM can naturally account for the alternation of long and short outbursts. There is no reason to believe that tidal forces are increased by more than one order of magnitude, even when the disc radius has shrunk well inside the 3:1 resonance radius, which is the essence of the TTI model. The simplest hypothesis one can make is that the tidal forces strongly increase beyond the resonance radius, but there need not be an hysteresis. The superhump phenomenon seems, however, to be linked with the 3:1 resonance; it is nevertheless not a surprise that the 3:1 resonance radius is reached only during superoutbursts since the disc extension is maximum during these outbursts (see Fig. \ref{fig:su}).

One should finally note that both the DIM and the TTI model assume an axisymmetric disc, which is clearly not the case, at least during superoutbursts. 

\subsection{Outbursts ending a standstill}
\label{sec:zc}

IW And and V513 Cas are Z cam systems that have displayed an anomalous behavior: the standstill period was terminated by an outburst \citep{s11,sal13}. \citet{hl14} showed that outbursts of mass transfer (with a duration of a few days, with a short rise time and an exponential decay) from the stellar companion will account for the observed properties of V513 Cas and IW And, provided they are followed by a short but significant mass-transfer dip. The total mass involved in outbursts is of the order of $10^{23}$ g. 

\citet{k19} and \citet{kpp19} discovered 4 other systems exhibiting the same type of behavior, and concluded that they cannot be due to a mass transfer outburst because of the regularity of the light curve of FY Vul, one of the systems they observed. Since we do not understand the mechanism that generate these mass-transfer outbursts, the argument is weak. They  proposed as an alternative possibility that, in these anomalous systems, the disc radius grows during standstills until it reaches the 3:1 resonance limit, at which point an outburst is triggered. They also argue that the mass transfer rate is the same during standstills and during the outbursting period because the mean optical luminosity does not change during these periods, contrary to what is found in normal Z Cam systems. Their assertion about the normal Z Cam systems is true in general, but not for individual objects; \citet{hrt98} found that, for normal Z Cam stars, there is a large scatter of the difference between the standstill magnitude and the average outbursting phase magnitude between individual objects, with differences ranging between -0.42 to +0.50. Moreover, in the mass transfer outburst model, one expects that the luminosity during the outbursting phase, {\em including the initial mass transfer outburst} can be larger than during standstill. Finally, \citet{hl14} showed that, if one artificially increases the tidal torque for a few days while keeping the mass transfer rate constant, the tidal instability model does produce an outburst, but after that, the system returns into a standstill phase and does not enter a dwarf nova outburst phase.

\subsection{Stunted outbursts}
\label{sec:stunted}
A closely related topic is the case of stunted outbursts. These are weak (typically 0.4 -- 1 mag) outbursts with  time properties (recurrence, duration) similar to those of dwarf nova outbursts, but much smaller amplitudes \citep{hrt98b}. These outbursts are usually found in novalike systems, and their origin is unclear. It was suggested that they might be due to some instability in the accretion disc because they have, apart from their brightness, similar properties to dwarf nova outbursts \citep{h01}. If the instability is the same as for dwarf novae, the low amplitude must be due to the presence of a constant background light of unknown origin that exceeds by far the disc luminosity in quiescence. The possibilities mentioned by \citet{h01} are rotational energy, magnetic energy, radiation from a hot white dwarf, or hydrogen burning on the white dwarf. It is also possible that they are due to a different disc instability that remains to be discovered. Another possibility is that stunted outbursts are due to a mass-transfer instability. \citet{rhh18} found that UU Aqr showed clear evidence for hot spot enhancement during half of the stunted outbursts they observed; but the other half did not show any sign of an increase of the mass transfer rate. A variation of the hot spot luminosity can be due either to a variation of the mass transfer from the secondary, with a response time of order of the dynamical time, or, if the transfer rate is kept constant, to a variation of the outer disc radius on a time scale of order of the viscous time; here, the hot spot luminosity variations are large and cannot be easily accounted for by a change in the hot spot position. Moreover, the disc outer radius increases during an outburst, because of the outwards flow of angular momentum generated by the accretion burst, so that the hot spot luminosity is expected to decrease for a constant mass transfer rate.

Numerical simulations indicate \citep[see e.g.][]{hl14} that, in order to account for observed luminosity of the stunted outbursts, the mass transfer rate must increase by about a factor 10. The reason for such a large increase is unknown; flares are one possibility, but it is not clear that the required amplitude can be obtained. One may also wonder why these outbursts are found in systems that cover a wide range of orbital periods and hence secondary masses and evolutionary states. The systems discussed by \citet{h01} have orbital periods ranging from 3.9 hr (UU Aqr) to 1.48 day (X Ser). As already mentioned, from a more general point of view, the reasons and mechanisms for mass transfer variations in cataclysmic variables are poorly understood.

\subsection{Disc vertical structure}
The DIM is based on the standard disc model and is therefore assumed to be geometrically thin and laying in the orbital plane of the binary. This is an oversimplifying assumption that proves to be at best inaccurate in a number of cases. Strictly speaking, this is not a deficiency of the DIM itself, but of the standard disc model that might have important consequences on the predictions of the DIM.

Irradiation of the accretion disc is observed to be quite important in low mass X-ray binaries, and yet screening of the central source by the accretion disc itself should prevent heating of the outer disc; this lead \citet{dlh99} to postulate that the disc must be warped. This might also be the case for systems containing accreting white dwarfs, in particular if the white dwarf is hot \citep[see e.g.][for the case of symbiotic stars]{bhl18}. Many models have been proposed to illustrate how a warp could develop in an accretion disc \citep[see e.g.][to name a few]{l92,sm94,p96,l99,od01}. The abundance of possible explanations may be interpreted as the absence of a generally accepted model, although an instability due to radiation forces on the disc as suggested by \citet{p96} seems however to be a good candidate.  

Modulations of the light curve at a period slightly shorter than the orbital period are observed in a number of cataclysmic variables; these are the so-called negative superhump. These are believed originate from the retrograde precession of a tilted accretion disc \citep{bmm85}. SPH simulations show that the modulation of the optical light is due to variations of the position of the hot spot on the disc as a result of the tilt and of the precession \citep{wb07}. The origin of the tilt is largely unknown; as a difference to X-ray binaries, it cannot be due to radiation pressure on the disc; distortion of the magnetic fields from the secondary have often been invoked \citep{bow88}, even though the details are not well understood. Using SPH simulations, and based on a model by \citet{l14} in which the primary dipole is misaligned with its spin axis, \citet{tw15} on the other hand showed that the period deficit was in good agreement with observations.

It is also usually assumed in the DIM that matter is incorporated in the disc at its outer edge. There are however observational evidences that the stream from $L_1$ overflows the accretion disc \citep[see e.g.][]{l89,h93,hr94,s08}. This has several important consequences. First, the outbursts may be affected; \citet{sh98}, showed that when 25\% of the stream overflows the disc, the deposition of mass in the inner parts is sufficient to change the behavior of the heating and cooling fronts, and cause reflares, thereby changing significantly the shape of the eruption light curves (see Eq. \ref{eq:defT}). Note however that these calculations were performed using a 1D code, while the overdensity pattern is 2D, as shown by SPH simulations \citep{ksh01}. These simulations showed that, for large overflow fractions, the overflow pattern can be even more complex than assumed by \citet{sh98}; and this effect still needs to be properly taken into account in the DIM. The disc radius evolution is also modified in a non trivial way because there will be less mass in the outer disc and the tidal torque will be modified and this may be important if the mass transfer rate is modulated (see sect. \ref{sec:tti}). In the case of intermediate polars, it may also explain why accretion may occur both via the accretion disc and directly via the interaction of the overflowing stream with the magnetosphere \citep{h93}. Finally, this changes the hot spot luminosity and the way it depends on the mass transfer rate and on the outer disc radius.

These examples clearly show that the disc vertical structure is clearly more complex than assumed in all DN models.

\subsection{Spectral evolution} {\label{sec:spectra}
Finally, a detailed comparison between models and observations has to involve the calculation of the disc spectrum. To zeroth order, such a spectrum can be calculated by simply summing blackbodies, once the radial distribution of the effective temperature $T_{\rm eff} (r)$ is known. This is not a good approximation notably in the UV, where significant deviations from a blackbody are expected. A better approach is to sum stellar spectra \citep{w84}, but models still do not fit well the observations. As stellar spectra do not include the fact that the vertical gravity in a disc varies with height and that viscosity dissipates energy in the atmosphere, these effects were included in \citet{h90} \emph{TLUSDISK} model. But again, the spectral models cannot reproduce well the observed spectra of bright, steady  systems \citep[][and references therein]{np19}. These latter authors suggested that the problem might be with the assumptions on the angular momentum transport (see Sect. \ref{sec:viscosity}); it may be also that the impact of winds on accretion disc spectra is not well taken into account \citep{mkl15}, and in any case, one would like to couple a radiative transfer code to the MRI simulations, as the vertical distribution of energy release is important. The situation is even worse for cold disks, despite some progress \citep{ilh10}. As mentioned earlier, a number of global, 3D MHD numerical simulations have been conducted \citep[see e.g.][]{jsz16,jsz17}, but these, by far, do not include the thermodynamics required to describe the thermal instability and cannot be used to estimate the disc spectra. 

This may explain why the calculation of a spectrum as predicted by the DIM and its comparison with observations is not often done, and DIM based spectra are usually based on very simple and crude assumptions (summation of blackbodies or stellar spectra). It would be worth, however, to try to fit the color evolution of dwarf novae using the DIM, since it has been known for long that dwarf novae follow a loop in the $U-B$, $B-V$ plane \citep{b80}. This requires, however, to include all sources of light, in particular from the white dwarf, the hot spot and the (possibly irradiated) secondary that have a significant contribution during quiescence. This work is currently under way.

\subsection{Rebrightenings} \label{sec:rebr}
Rebrightenings are sometime observed at the end of long outbursts of SU UMa stars, in particular of the WZ Sge subtype that only exhibit superoutbursts, as well as at the end of soft X-ray transients. There are instances where rebrightenings appears as a short, well separated outburst after the main outburst; this is the case for example of V585 Lyr as observed by \emph{Kepler} \citep{ko13}. In other systems, rebrightenings occur during the decline from maximum. In the latter case, it could in principle be explained by a reflection of the cooling wave that propagates from the outer disc edge. \citet{dhl01} showed the light curves produced this way do not in general resemble the observed ones; but \citet{hl16} were able to reproduce a rebrightening occurring during the decline of an outburst in the case of an ultracompact X-ray binary. Reflection of the cooling wave due to stream overflow \citep{sh98} or a modification of the $\alpha$ prescription suggested by \citet{ckb16} could also be an option, but these have not been explored much and the comparison of the predicted and observed light curves does not look very promising: they tend to occur much earlier during the decline than observed.

\begin{table*} 
\caption{Additional ingredients to the DIM for various classes of transient systems considered here.}
\label{tbl:prop}           
\begin{tabular}{lccccc}                   
\hline                    
                         &  U Gem  & Z Cam & SU UMa & WZ Sge & SXT  \\
\hline                    
Inner disc truncation    &    ?    &   ?   &   ?    &   +    &  ++  \\
Disc irradiation by WD   &    +    &   +   &   +    &   +    &  -   \\
Self irradiation of the disc & -   &   -   &   -    &   -    &  ++  \\
Secondary irradiation    &    +    &   +   &  +/++  &   +/++ &  +   \\
Disc winds               &   ++ ?  &  ++ ? &  ++ ?  & ++ ?   &  ++ ?\\
Tidal instability        &    -    &   -   &   +/++ &  +/++  &   ?  \\
\hline                                  
\end{tabular} 
\end{table*}

The case of well separated rebrightenings is even more challenging. \citet{bh02} suggested that these rebrightenings could be normal short outbursts occurring after a superoutburst while the mass transfer rate was still higher than normal as a result of irradiation of the secondary during the superoutburst; however, the intervals they found between the main outburst and the reflares were twice as long as observed. \citet{omm01} proposed that reflares could be due to the viscosity remaining high after the superoutburst and decaying exponentially with time; they were able to reproduce the light curve of RG Cnc that exhibited six rebrightenings after its 1996 -- 1997 outburst, but the  calculations were made using a very crude modelling of the accretion disc. \citet{mm15} elaborated on this possibility,  but as they did not calculate light curves, it is difficult to ascertain it. 

\emph{Kepler} also observed mini-rebrightenings in V585 Lyr, between the superoutburst and a 3 mag. rebrightening that occurred 6 days after a sharp 3 mag. drop in the lightcurve terminating the superoutburst \citep{ko13}. Their amplitude was about 0.4 -- 0.5 mag., with a periodicity of half a day. \citet{mm15} suggested that these are due to the small wiggles in the S curve at temperatures of order of 3000 -- 4000 K. Their model requires however that $\alpha_{\rm c} = 0.2$, which they justify by stating, as mentioned above, that the superoutburst has modified the viscosity and that this modification persists for a while. However, with such a high value of $\alpha_{\rm c}$, the cooling front ending the superoutburst would not have been able to propagate far in the disc, which means that these mini-rebrightenings would occur while the system is still very bright, contrary to what is observed.

Soft X-ray transients also frequently show rebrightenings, but despite the similarities with CVs noted by \citet{khv96}, the mechanisms could be different as noted by \citet{l01}.

\section{Conclusion}
Table 1 summarizes the ingredients that must be added to the DIM for each subclass of transient sources. A "++" indicates that the ingredient is major and is requested to account for the whole subclass; a "+" that it is an important one that may be needed to account for some specificities (e.g. the irradiation of the secondary for long outbursts in U Gem); a "-" that it is not important, although it may be present. In the case of competing models, both options are shown.

As can be seen, many additional ingredients are often needed in addition to the pure DIM model to account for the observations. With these, the DIM, despite its many physical limitations, has been very successful in explaining the outbursts of dwarf novae and, to a lesser extent, of soft X-ray transients. These ingredients, however, have free parameters that are weakly constrained, which makes it difficult to fully test the DIM. It is for example crucial to understand the response of the secondary star to irradiation and to estimate the resulting increase of the mass transfer rate. At present, this increase is either ignored, or parametrized in a very crude way. Similarly,  intrinsic fluctuations of the mass transfer from the secondary are observed, have a large amplitude (a factor 10 in AM Her systems), but are not understood. 

The DIM itself suffers a number of weaknesses and deficiencies that have been noted in this review. The most notable is our limited understanding of the processes that transport angular momentum. While there is a general consensus that MRI is the mechanism transporting angular momentum in bright accretion discs, the situation is far from being clear in the low state. Even in the high state, it is unclear how realistic shearing box simulations are. Ideally, one would like to solve the full 3D MHD structure including radiation, but this is a formidable task.

Finally, in its current version, the DIM is a 1D model. Taking properly into account tidal torques in 2D models, or the impact of the stream on the disc, or its possible overflow, would be highly desirable. 

In summary, future progresses will come from a better understanding of the physical processes occurring in accretion discs, and in particular angular momentum transport and the inclusion of winds (and the torque they exert on the disc itself), an accurate modelling of the surface and subsurface layers of the secondary to account for the mass transfer fluctuations, and, to a lesser extent, an improvement of the numerical codes with the inclusion of 2D effects, at least until full 3D MHD numerical simulations will become available. In the meantime, the DIM will remain a major tool for modelling the outbursts of dwarf novae and soft X-ray transients, because it incorporates the thermodynamics in a realistic way.

\section{Acknowledgement}
I thank Jean-Pierre Lasota for helpful discussions and a careful reading of this manuscript. I am also grateful to the referees for their detailed and thorough comments that where very helpful for improving this paper.

% Loading bibliography database
\bibliographystyle{model2-names}
\bibliography{refs}

\end{document}